\documentclass[pdflatex,sn-mathphys-num]{sn-jnl}

\usepackage{graphicx}%
\usepackage{multirow}%
\usepackage{amsmath,amssymb,amsfonts}%
\usepackage{amsthm}%
\usepackage{mathrsfs}%
\usepackage[title]{appendix}%
\usepackage{xcolor}%
\usepackage{textcomp}%
\usepackage{manyfoot}%
\usepackage{booktabs}%
\usepackage{algorithm}%
\usepackage{algorithmicx}%
\usepackage{algpseudocode}%
\usepackage{listings}%
\usepackage{comment}%

\usepackage{color}

\usepackage[numbers,sort&compress]{natbib}
\newcommand{\xfnm}[1][]{\ifx!#1!\else\unskip,\space#1\fi}

\begin{document}

\title[Article Title]{Optical Ground Station Diversity for Satellite Quantum Key Distribution in Ireland}


\author*[1]{\fnm{Naga Lakshmi} \sur{Anipeddi}}\email{nagalakshmi.anipeddi@waltoninstitute.ie}

\author[1]{\fnm{Jerry} \sur{Horgan}}\email{jerry.horgan@waltoninstitute.ie}

\author[1,2]{\fnm{Daniel K L} \sur{Oi}}\email{daniel.oi@strath.ac.uk}

\author[1]{\fnm{Deirdre} \sur{Kilbane}}\email{deirdre.kilbane@waltoninstitute.ie}

\affil[1]{\orgdiv{Walton Institute}, \orgname{South East Technological University}, \orgaddress{\city{Waterford}, \postcode{X91 P20H},  \country{Ireland}}}

\affil[2]{\orgdiv{SUPA Department of Physics}, \orgname{University of Strathclyde}, \orgaddress{\city{Glasgow}, \postcode{G4 0NG},  \country{United Kingdom}}}



\abstract{Space quantum communications is a potential means for establishing global secure communications and quantum networking. Despite pioneering demonstrations of satellite quantum key distribution, considerable challenges remain for wide deployment such as the local effects of the atmosphere on the transmission of single-photon level quantum signals. As part of Ireland's efforts to establish quantum links with the rest of Europe and further afield, we present a preliminary study of the feasibility of satellite quantum key distribution taking into account geographic and weather effects on the space-Earth channel. Weather data over 5 years covering 4 locations across Ireland were used to assess performance and the prospects of optical ground station (OGS) geographic diversity to improve service availability. Despite significant cloud cover that may reduce the performance of a single OGS location, the use of a 4-OGS network can provide up to 45\% improvement for a single satellite exploiting anti-correlation in cloud cover, though most gains are achieved with 2 or 3 OGSs.}




\maketitle

\section{Introduction}\label{sec:Introduction}

Quantum Key Distribution (QKD)~\cite{sidhu2021advances,Pirandola:20, RevModPhys.74.145, RevModPhys.81.1301, diamanti2016practical, BENNETT20147} has the potential to provide secure information exchange between any two locations on Earth. Fundamental tests and advanced QKD protocols in the field of quantum information science have paved the way for quantum communications (QC)~\cite{doi:10.1063/1.5023340, Lucamarini2018}. Several optical fibre based terrestrial quantum networks have been deployed in cities such as Bristol, Cambridge, Madrid, and Vienna~\cite{wang2023field,Dynes2019,Hbel2023DeployedQN,35677c78eaba42f5bb61081e34134b2a}. However, in the absence of quantum repeaters, the practical range for direct transmission is limited to a few hundred kilometers due to the exponential losses in fiber, making it impossible to attain global range over large distances~\cite{sidhu2021advances,PhysRevLett.117.190501, PhysRevLett.124.070501, Fang2020, RevModPhys.94.035001}. To an extent quantum repeaters may bypass the direct transmission constraint, but their performance requirements make them unsuitable for growing to the intercontinental ranges required for global scale-up of QC~\cite{Sidhu2023}. Satellite quantum key distribution (SatQKD) offers the possibility to overcome 
this distance limitation and the Micius satellite has demonstrated the feasibility of the concept, as part of a China-wide 
trusted-node heterogeneous QKD network and intercontinental links~\cite{Chen2021,sidhu2021advances,PhysRevLett.120.030501}. The implementation of global-scale SatQKD will be a precursor to the quantum internet~\cite{sidhu2021advances}.

Using a satellite as a node can provide long-distance links between national sub-networks and enable international connection diversity~\cite{sidhu2021advances,pirandola2021limits}. In the case of Ireland, the opportunity to collaborate with international QC projects and smallsat missions, such as SpeQtre (UK-Singapore)~\cite{Speqtre}, SpeQtral-1 (Singapore)~\cite{SpeQtral}, QUBE-1~\cite{Knips:22}, QUBE-II~\cite{dlr190981}, and QuNET (Germany)~\cite{QuNET}, QEYSSat (Canada)~\cite{10.1117/12.2041693} will provide the opportunity to demonstrate such capability. Additionally, Ireland can engage with initiatives like the Satellite Platform for Optical Quantum Communications (SPOQC) funded by the UK Quantum Communications Hub~\cite{zhang2023end}, the European Quantum Communication Infrastructure (EuroQCI) funded by the European Commission and European Space Agency~\cite{EuroQCI}, and missions launched by the European Space Agency, like Versatile Optical Lab for Telecommunications (OPS-SAT VOLT)~\cite{OPSAT}, Infrastructure for Resilience, Interconnectivity and Security by Satellite (IRIS$^{2}$)~\cite{orsucci2024assessment}, Security And cryptoGrAphic mission (SAGA)~\cite{riccardi2023space}, EAGLE-1~\cite{orsucci2024assessment} and future quantum technology missions.

A single satellite functioning as a trusted QKD node or to distribute entanglement can replace a chain of nodes~\cite{Liao2017, PhysRevLett.120.030501, Ren2017, Villar:20, Sidhu2023}. Practical implementation of SatQKD drives miniaturization of space-based quantum sources and optical systems to align with optical ground stations (OGSs) resource requirements and satellite technology~\cite{Oi2017,PhysRevApplied.5.054022,Jennewein_2013,Kerstel2018}. To maximize system performance, we can optimize downlink scheduling taking into account network demands and site conditions~\cite{Polnik2020,Wang_2023,Liao2017}. A significant challenge for low-Earth orbit (LEO) satellites is the restricted OGS overpass time that limits the number of received signals together with the highly variable channel loss~\cite{Sidhu2023}. This causes significant finite block-size effects that constrains the volume of secure keys that can be generated in a single overpass~\cite{Sidhu2022}. For precise estimates of SatQKD performance, site-specific data, e.g. detailed background light levels, together with system-level parameters such as QKD protocols and sub-system characteristics are required for high-fidelity modelling~\cite{10487878}. However, for preliminary feasibility analysis, we can estimate the potential for SatQKD using a more general approach.

Here, we perform an initial assessment for SatQKD links for Ireland, taking into account weather statistics and possible mitigation using OGS geographic diversity. In this work we assume modest space and ground segment facilities that are adaptable and configurable for SatQKD implementation. Several research teams around the world are moving towards distributing quantum keys with satellites and LEO missions with expected launches as early as 2025-26~\cite{progress,sidhu2021advances}. We account for effects due to the atmospheric channel, wavelength, location, size of transceiver, time and range of operation, angle, and altitude, along with satellite overpass time at an optical ground station. The upper bound of the expected key capacity for four different locations in Ireland is modelled for a single satellite in sun synchronous orbit. Our analysis is protocol agnostic and leaves open the question of implementation and other specific factors that will influence actual expected key volume. As our main concern is with site selection, it is the relative performance that is of interest. The paper is structured as follows: we explore the configuration of a satellite communication system and model various signal losses to optimize system performance, and we evaluate the expected secret key capacity using a satellite passing over Ireland under cloud cover.

\section{Space to Ground Optical Channel}\label{sec:Space To Ground Channel}

\begin{figure}[htbp]
\centering
\includegraphics[width=\textwidth]{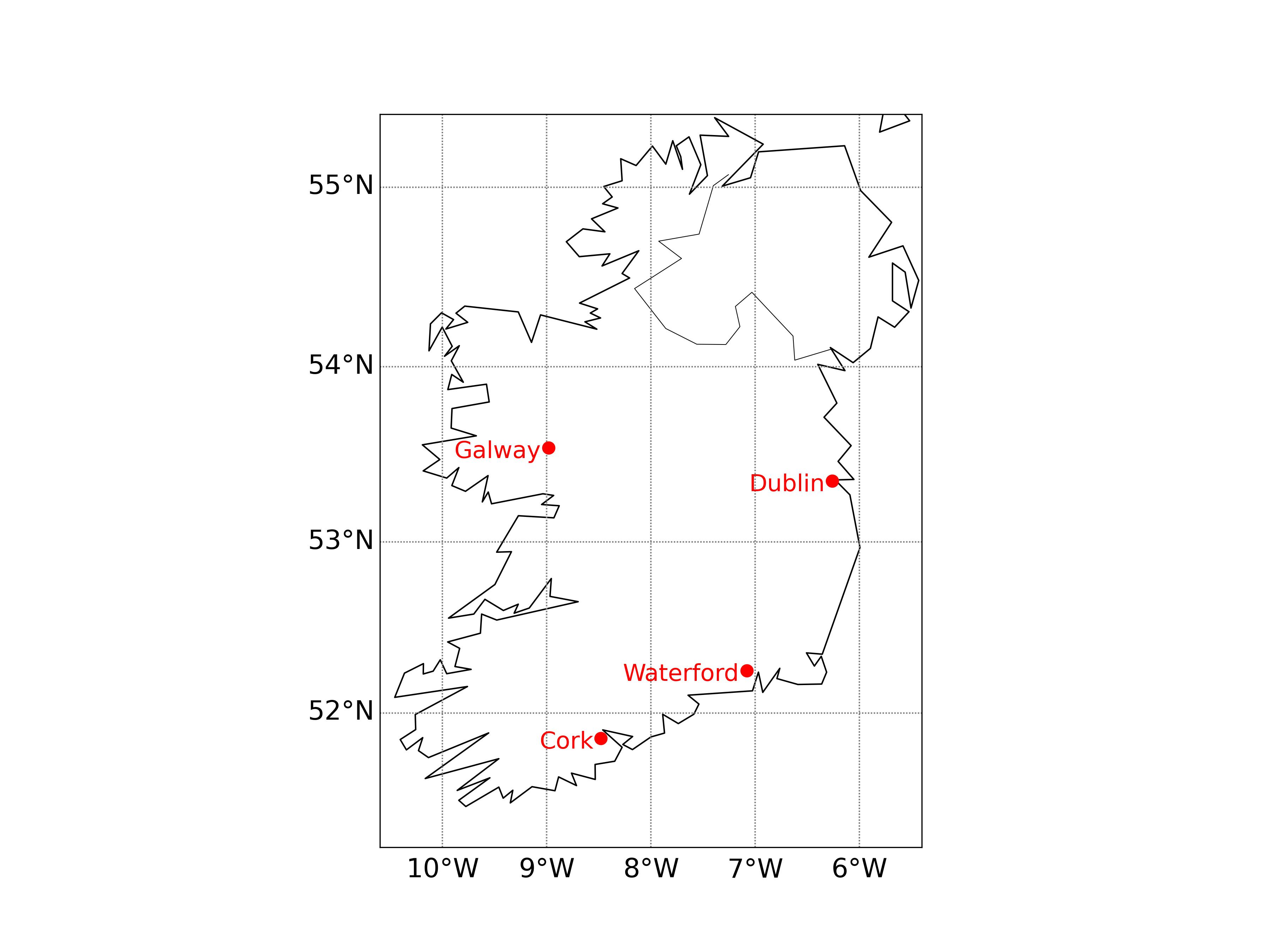}     
\caption{Potential Ireland OGS sites. Four locations in Ireland's coastal regions (latitude, longitude), Dublin $(53.35^\circ, -6.25^\circ)$, Waterford $(52.25^\circ, -7.08^\circ)$, Cork $(51.85^\circ, -8.48^\circ)$ and Galway $(53.54^\circ, -8.98^\circ)$ have been considered for siting OGSs. These regional centres are with high population density in urban locations that would benefit from SatQKD connections to local fibre networks.}
\label{fig:locations}
\end{figure}

We consider 4 locations distributed across Ireland shown in Fig.~\ref{fig:locations}, Dublin, Waterford, Cork, and Galway, as they are among the top five urban populated areas lining the Irish coastal regions, strategically positioned to facilitate connectivity with other countries. Their robust infrastructure and geographical location make them vital hubs for international communication networks, fostering seamless connectivity by bridging the national network internally. In this section we analyse the satellite orbital dynamics to determine coverage and model the channel performance for the downlink from the satellite to the ground stations in Ireland.

\subsection{Satellite to Ground Channel Model Analysis}\label{sec:Satellite to Ground Station Communication Channel Analysis}

We will characterise the space-ground channel primarily by the end-to-end loss $\eta_{\lambda}(\theta)$ experienced when sending single-photon level signals, which depends on the wavelength $\lambda$ of the photons and the angle $\theta$ of the line-of sight between the satellite and OGS. For a satellite in a circular orbit, the range and $\theta$ are in one-to-one correspondence. The link efficiency has several contributions: atmospheric extinction $\eta_{atm}$, diffraction $\eta_{diff}$, and a lumped estimate for other losses $\eta_{other}$ such as turbulence, pointing error, optical and detection efficiencies,
\begin{equation}
\eta_{\lambda}(\theta) = \eta_{atm}(\lambda,\theta) + \eta_{diff}(\lambda,\theta) + \eta_{other},
\label{eqn:totalloss} 
\end{equation}
where we express all losses in dB. We develop a channel model to analyse the performance of the communication link between the satellite and the ground station. Effects included are: 1. Satellite to ground station geometry; 2. Atmospheric extinction; 3. Diffraction; and 4. Other lumped losses. A more detailed treatment of turbulence effects has been relegated to future study.

\subsubsection{Satellite to Ground Station Geometry}\label{sec:Satellite to Ground Station Geometry}

\begin{figure}[htbp]
\centering
\includegraphics[width=\textwidth]{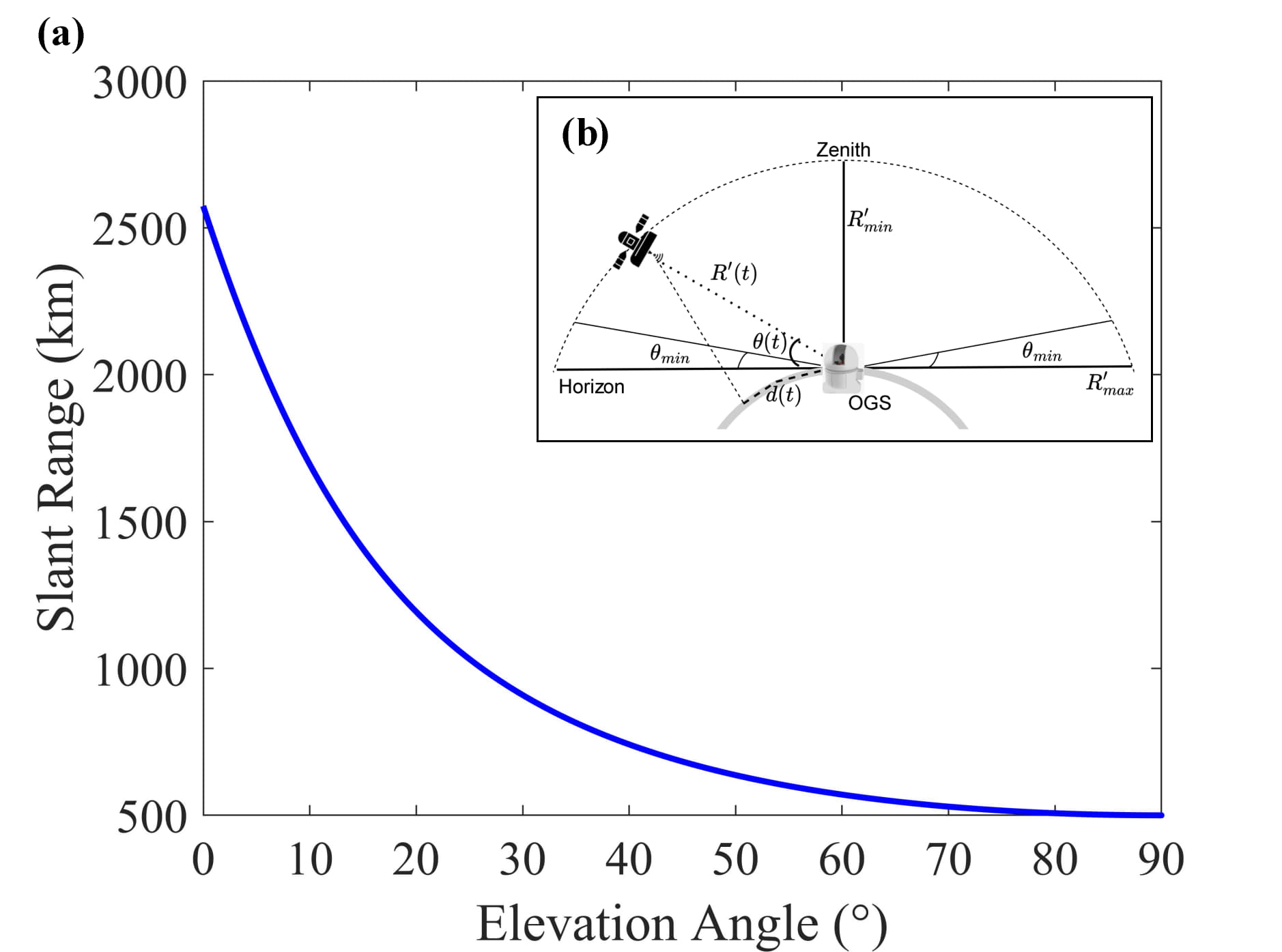}          %
\caption{Satellite-OGS geometry and slant range. (a) The dependency of slant range $R'$ on elevation angle $\theta$ for a satellite orbiting the Earth at an altitude $h=R'_{min}$ of $500~km$. (b) Schematic diagram of the geometry of a satellite in zenith overpass and communicating with an optical ground station, covering ground track distance $d$, in transmission time $t$, with minimum elevation $\theta_{min}$. 
}
\label{fig:rangeandgeometry}
\end{figure}

The basic geometry of a satellite orbiting the Earth and establishing a communication link with an optical ground station (OGS) is shown in Fig.~$\ref{fig:rangeandgeometry}$(a). The slant range $R'$ represents the distance between the satellite and the ground station,
\begin{equation}
R' = \sqrt{R_e^2 + (R_e + h)^2 - 2R_e(R_e + h)\cos(\theta)},
\label{eqn:range}
\end{equation}
where $R_e$ is the radius of the Earth, $h$ is the altitude of the satellite, and $\theta$ is the elevation angle between the ground station and the satellite. The dependency of the slant range on the elevation angle is shown in Fig.~$\ref{fig:rangeandgeometry}$(b) for a satellite located at an altitude $h = 500km$. At an elevation angle of $\theta = 90^\circ$, the satellite is directly above the ground station and the slant range is a minimum resulting in maximum transmission of the signal between the satellite and ground station. As the elevation angle decreases, the distance between the satellite and the ground station increases, leading to a greater loss of signal. This effect is more pronounced at elevations less than $30^\circ$, and establishing a reliable signal becomes challenging.

\begin{figure}[htbp]
\centering
\includegraphics[width=\textwidth]{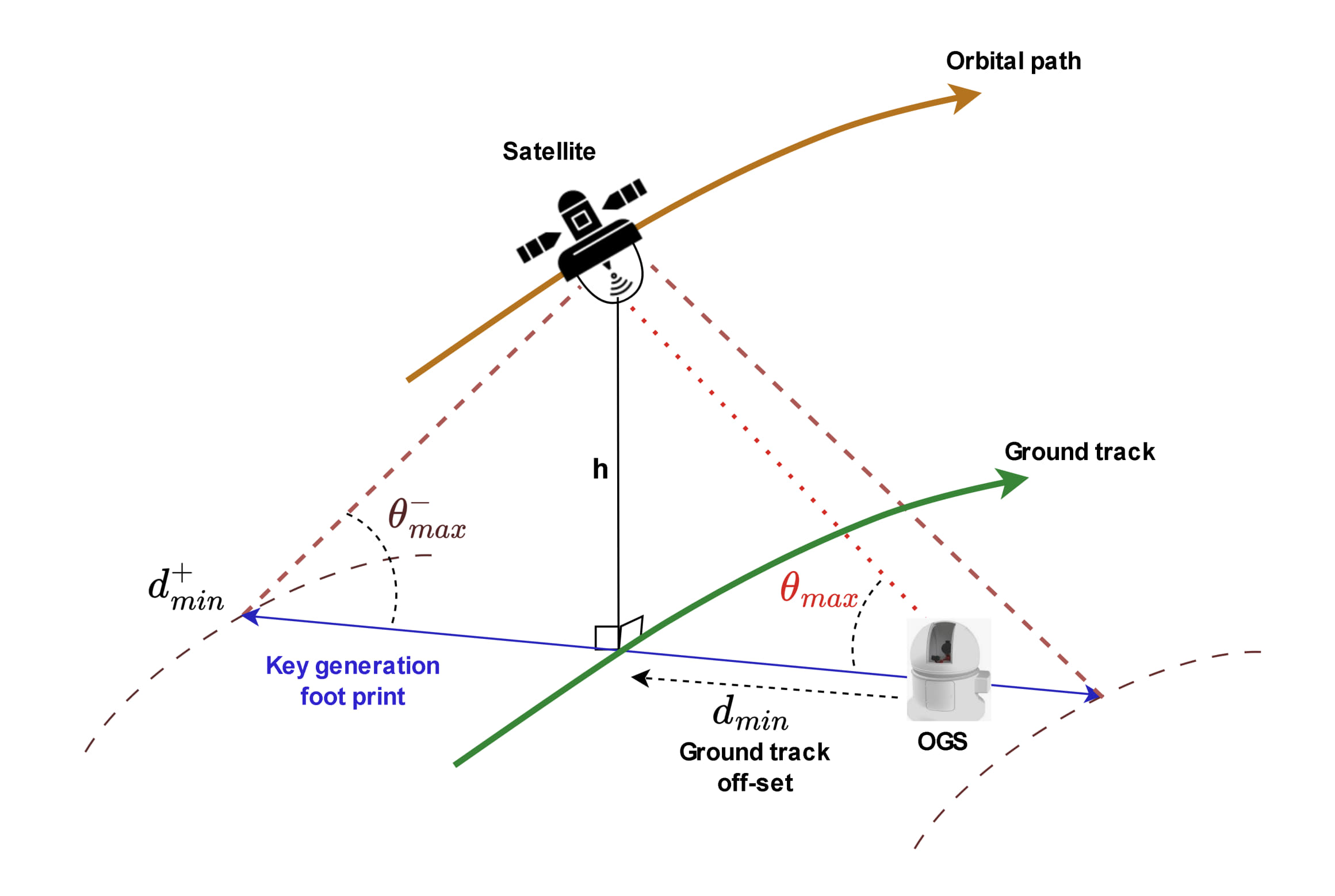}
\caption{General overpass geometry of the satellite with smallest elevation angle $\theta_{max}$ that allows for the generation of a secure key is referred to as  $\theta^{-}_{max}$ and defines the operational footprint for SatQKD, which spans a distance of $2d^{+}_{min}$ (adapted from~\cite{Sidhu2022})}
\label{fig:generaloverpassgeometry}
\end{figure}

In general, a satellite may pass near an OGS, though not necessarily directly overhead, as shown in Fig.~\ref{fig:generaloverpassgeometry}. The ground track of the orbit approaches the OGS with a minimum ground track offset of $d_{min}$ resulting in the satellite-OGS line-of-sight reaching a maximum elevation $\theta_{max}$. Only in the special case of $d_{min}=0~m$ does the satellite pass directly over zenith ($\theta_{max}=90^\circ$).

\subsubsection{Satellite Orbital Overpass}\label{sec:Satellite Orbital Overpass}

The precise orbit, timing, and duration of satellite overpasses are vital for planning windows to optimize scheduling in the quantum communication network. The orbital period $T$ for the satellite to orbit the Earth is,
\begin{equation}
T = 2 \pi \sqrt{\frac{(R_e + h)^3}{G M}}
\label{eqn:timeperiod}
\end{equation}
The dependence of elevation and range as a function of time varies for different satellite overpass geometries and ground track offsets, $d_{min}$, and maximum satellite overpass elevations, $\theta_{max}$~\cite{Sidhu2023}. 

\begin{figure}[htbp]
\centering
\includegraphics[width=\textwidth] {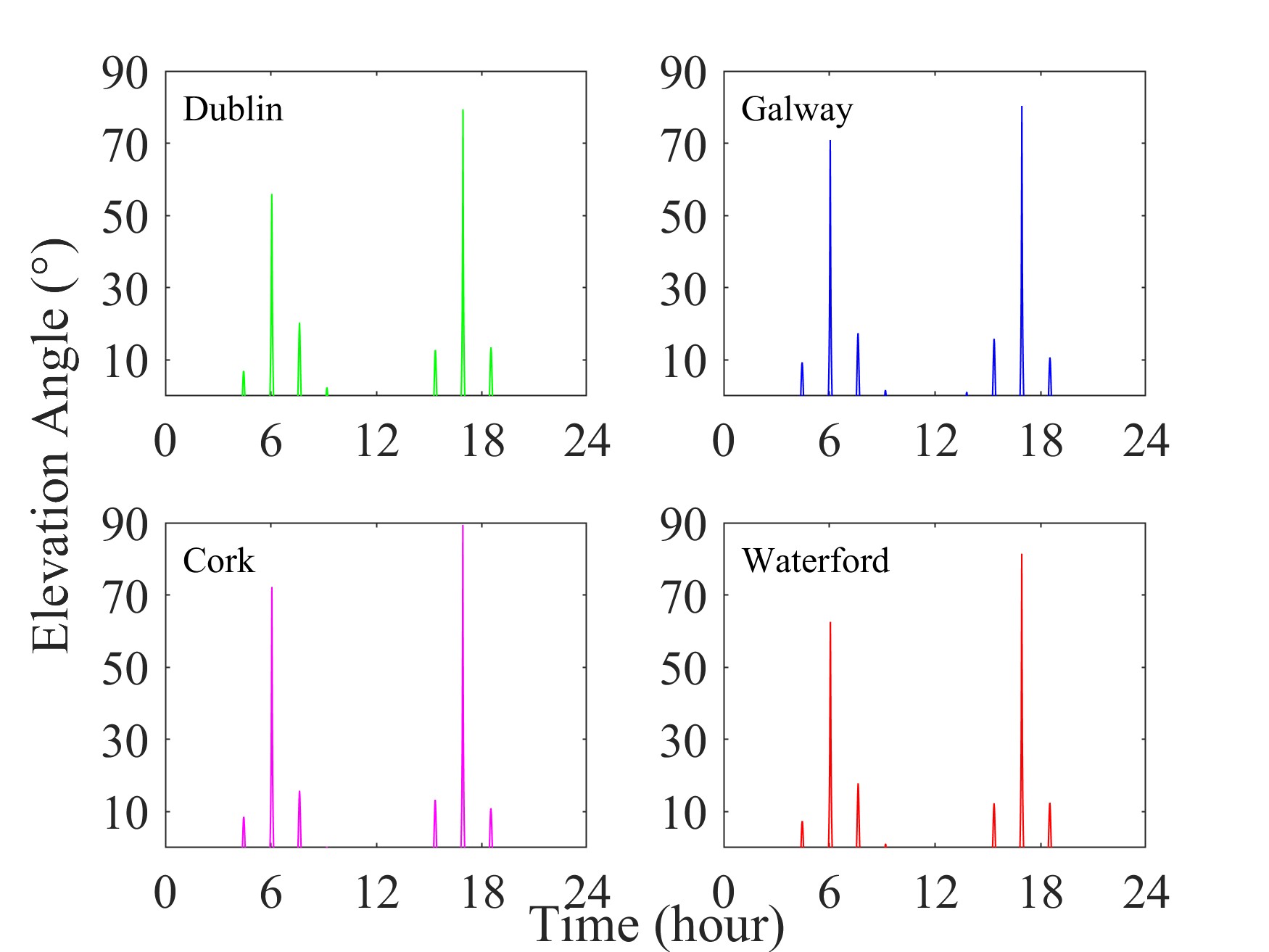}  %
\caption{STARLINK-3005 overpass simulations. Typically, a LEO dawn-dusk sun synchronous orbit gives 1 good overpass ($\theta_{max} \approx 90^\circ$) with 2 lower $\theta_{max}$ elevation passes as the Earth rotates under the orbit. A SSO has overpasses at the same local time, passes over 24 hours are shown here. The maximum elevation angles observed at Dublin, Waterford, Galway, and Cork are $79.40^\circ$, $81.43^\circ$, $80.36^\circ$ and $89.40^\circ$ respectively in this simulation run and occur at local times 06:00 and 17:00 with time windows of just over 12 minutes each.}
\label{fig:starlinkpasses}
\end{figure}

\begin{figure}[htbp]
\centering
\includegraphics[width=\textwidth] {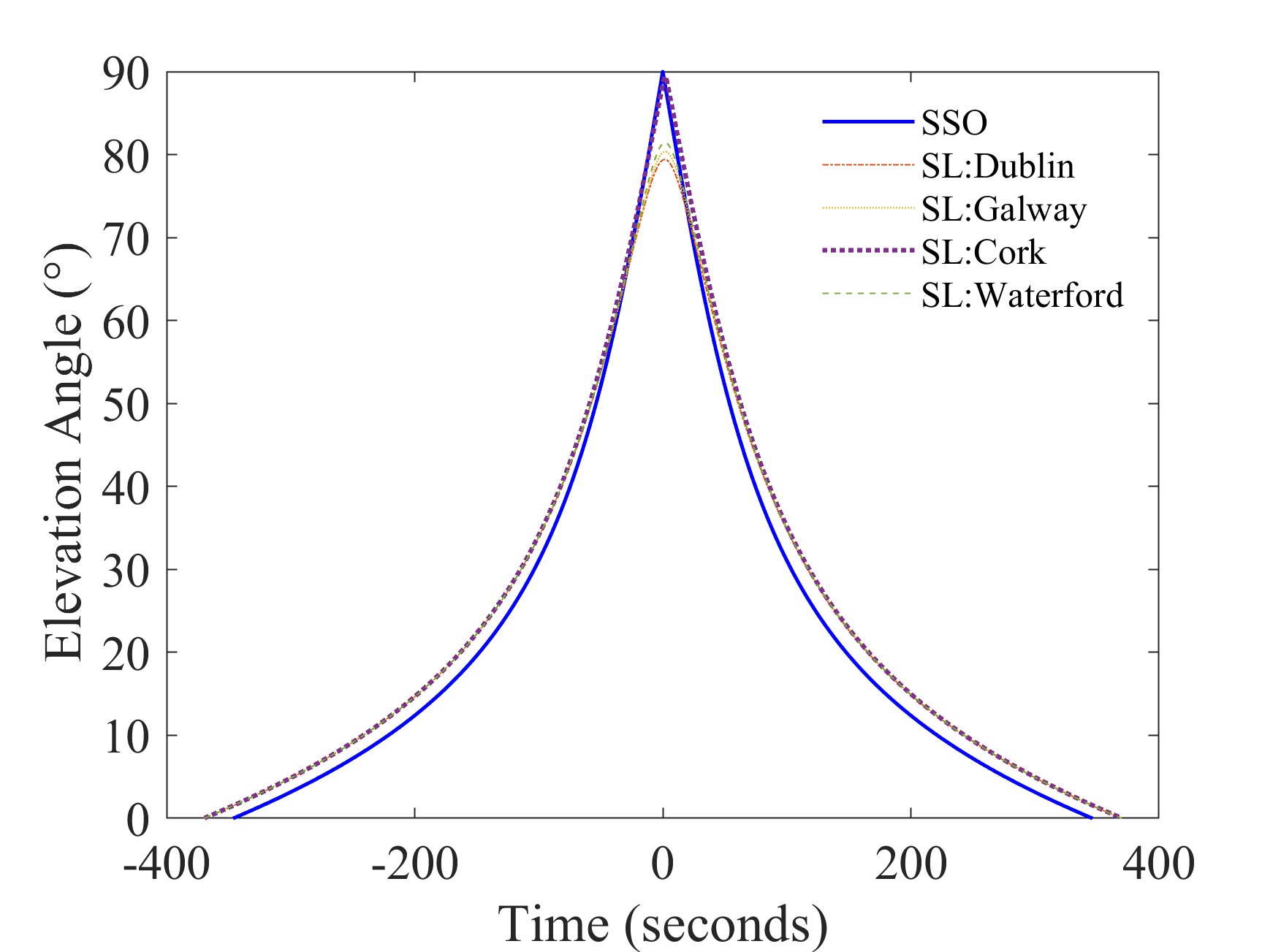}   %
\caption{Variation of angle of elevation with satellite overpass time for STARLINK-3005 ($h=564km$).
SL refers to STARLINK-3005 satellite passing over : Waterford, Dublin, Galway and Cork for $\theta_{max}$ as observed in Fig. \ref{fig:starlinkpasses} at around 17:00 hours. For comparison, SSO denotes a zenith overpass of a satellite at an altitude of 500km.}
\label{fig:starlinkpasscomparison}
\end{figure}

We use the Orekit open source flight dynamics software to model overpasses~\cite{Orekit}. We analyse sun synchronous orbits (SSOs) that ensure that the satellite passes over a location at the same local solar time. Typically, SatQKD missions choose a noon-midnight SSO so that at least one pass per day occurs with minimal background light (around midnight) and this is what we assume for our model. An example using Orekit is shown in Fig.~\ref{fig:starlinkpasses} for simulated overpasses of STARLINK-3005 ($97.7^\circ$ inclination at an altitude of 564 km) for a 24-hour period on 2023-11-27 using two-line element (TLE) data~\cite{TLEfile}. Though STARLINK-3005 is in a dawn-dusk SSO at an altitude of $h=564~km$, the overpass timing should be representative of other SSOs, albeit with a time shift. A comparison of these STARLINK-3005 overpasses with that of an SSO zenith overpass $h=500~km$ is shown in Fig.~\ref{fig:starlinkpasscomparison}.

\subsubsection{Atmospheric Extinction}\label{sec:Atmospheric Extinction}

In free space communication, atmospheric absorption and scattering degrades the signal and can be estimated using the Beer-Lambert equation~\cite{app122110944,NASAdoc},
\begin{equation}
    \eta_{atm}(\lambda)=\exp\left[-sec(\alpha)\int_{0}^{H}\gamma(\lambda,h)dh\right],
    \label{eqn:beer-lambert}
\end{equation}
where $\gamma$ is the atmospheric attenuation coefficient, $H$ is atmospheric channel vertical height, $\lambda$ is the wavelength, and $\alpha=90^\circ-\theta$ is the zenith angle. The extinction coefficient $\gamma$ depends on the integrated distribution of atmospheric constituents along the propagating path in addition to $\lambda$. We use the MODerate resolution atmospheric TRANsmission (MODTRAN) software, used extensively to simulate atmospheric extinction~\cite{10.1117/12.2050433,PhysRevApplied.16.044027,PhysRevA.99.053830} with the atmospheric generator toolkit (AGT) for site specific atmospheric models~\cite{Berk2016NextGM}.

\begin{table}[htbp]
\caption{MODTRAN settings for simulating transmittance and total radiance in the infrared region of the spectrum}
\label{table:MODTRANsettings}%
\begin{tabular}{@{}llll@{}}
\toprule
 MODTRAN Parameter& Value \\
\midrule
Radiative Transfer Option & Band model  \\
Band Model Resolution & 1 $cm^{-1}$  \\
RT Run mode & Solar and thermal \\
Multiple Scattering  & DISORT MS \\
Atmosphere Model  & Mid-latitude winter \\
Wavelength & $400~nm$ to $1600~nm$ \\
FWHM & $2~nm$ \\
Increment step size & $1~nm$  \\
Visibility Range   & $50~km$ \\
Geometric path  & Zenith\\
\botrule
\end{tabular}
\end{table}

\begin{figure}[htbp]
\centering
\includegraphics[width=\textwidth] {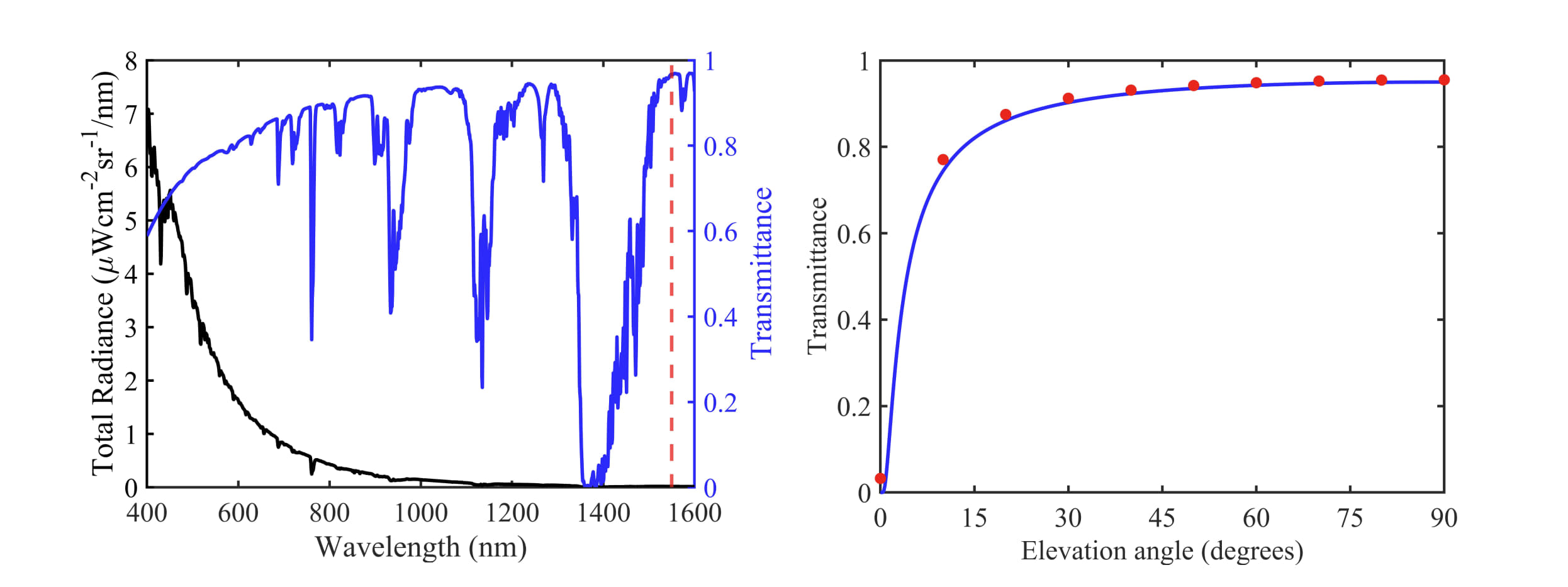}
\caption{MODTRAN results. (a) Transmittance and total radiance with wavelength in the visible and infrared spectral regions. The vertical red dashed line indicates 1550 nm. (b) Dependency of transmittance on elevation angle for 1550 nm (MODTRAN red dots, curve~Eq.\ref{eqn:transmittancewrtelevation}). For ease of comparison the four OGS location altitudes are assumed to be same and at sea level.}
\label{fig:transmittance}
\end{figure}

Fig.~\ref{fig:transmittance}(a) shows the simulated transmittance and total radiance for the spectral regions typically used for optical communications (MODTRAN settings in Table~\ref{table:MODTRANsettings}). There are high transmittance bands spread out over the visible and infrared regions. The low-attenuation near-infrared windows are supported by cost-effective, readily available single photon silicon detectors. However, the short-wave infrared (e.g. telecom-C) band may offer advantages for day-time QKD links due to reduced solar background, though at the cost of greater diffraction and the requirement for superconducting detectors.  
As there is no clear-cut choice of wavelength~\cite{orsucci2024assessment}, here we choose $1550~nm$ to align with potential advances in silicon integrated photonics and exploitation of existing telecomms components~\cite{Avesani2021}. The variation of atmospheric transmittance, $\tau_{atm}$, with the angle of elevation $\theta$ is given by,
\begin{equation}
\tau_{atm} = \tau^{sec(90^\circ-\theta)}. 
\label{eqn:transmittancewrtelevation}
\end{equation}
For simplicity we consider uniform atmospheric layers forming a slab atmosphere~\cite{atmosphere} which is a good approximation over the typical elevations for SatQKD. For instance, assuming a zenith $95\%$ transmission for $1550~nm$ as in Fig.~\ref{fig:transmittance}(a), the transmittance as a function of angle is shown in Fig.~\ref{fig:transmittance}(b). For each of the 4 locations the $1550nm$ transmittance is $\geq 0.9$, therefore we use $\tau_{atm}=0.9$ in further analyses.

\subsubsection{Diffraction Effects}\label{sec:Diffraction Effects}

Diffraction causes the transmitted beam to expand during propagation and we approximate the half divergence angle as,
\begin{equation}
\omega_{div} = 1.22 \cdot \left(\frac{\lambda}{D_{T}}\right),
\label{eqn:divergence}
\end{equation}
assuming a flat-top intensity profile over an unobstructed transmitter circular aperture of diameter $D_T$, where the far field intensity takes on an Airy profile~\cite{10.1088/2053-2571/ab2231}. The diffraction dB loss is approximated as~\cite{Polnik2020},
\begin{equation}
\eta_{diff} = -20 \cdot \log_{10} \left( \frac{D_{R}}{{D_{T}+(\omega_{div} \cdot R')}} \right),
\label{eqn:diffraction}
\end{equation}
where $D_R$ is the receiver diameter, after propagation across a slant range $R'$.


In a downlink scenario the laser beam travels from the satellite to the ground station and undergoes atmospheric perturbation only at the end of its path. As a result, most of the beam propagation occurs in vacuum, where the beam maintains its diffraction limit properties, while
the turbulent atmosphere is encountered only during the last $20~km$ of its path. On the contrary, for uplink the wavefront is distorted at the beginning of its path, resulting in greater loss of the signal during the beam propagation~\cite{Dequal2021}. Downlink is therefore more favorable for key generation than the uplink configuration.

\subsubsection{Other Losses}\label{sec:Other Losses}

Within our loss budget, we lump several other losses together as $\eta_{other} = \eta_{turb} + \eta_{point} + \eta_{opt} + \eta_{det}$. The total electro-optical inefficiency $\eta_{opt}$ of the OGS system is from different components: photon detection efficiency Si-SPAD, quantum receiver optics, collection telescope, interface and adaptive/tip-tilt optics between telescope and quantum receiver. We conservatively assign a fixed loss of $\eta_{opt}+\eta_{det}=12~dB$.
Turbulence and pointing errors lead to added loss in beam propagation to which we assign a fixed and conservative
value of $\eta_{turb}+\eta_{point}=8~dB$ independent of elevation/time. In the absence of on-site turbulence data and actual pointing/beam-steering performance observations, our simplified loss budget model provides comparative performance estimation across sites for preliminary analysis. Hence we assume that $\eta_{other} = 20~dB$.

\begin{figure}[htbp]
\centering
\includegraphics[width=\textwidth] {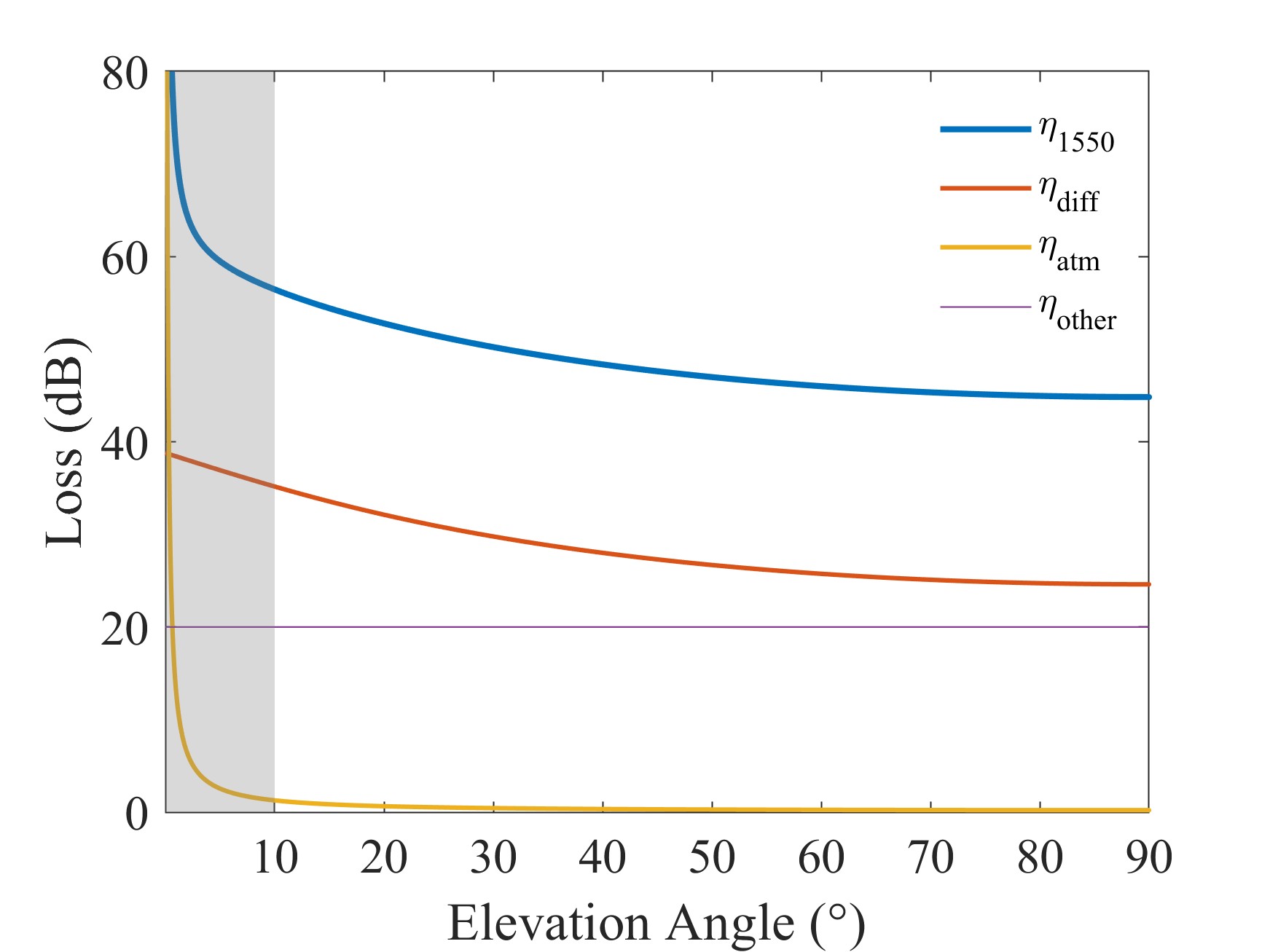}  
\caption{Dependency of total channel loss with angle of elevation. Both $\eta_{diff}$ and $\eta_{atm}$ vary with elevation and increase with decreasing elevations. We have lumped together several loss terms into $\eta_{other}$ and assumed a conservative fixed value of 20 dB. The grey shaded region indicates angles below $10^\circ$ which are not suitable for satellite communication due to the high losses. The zenith loss is $\eta_{1550nm}(90^\circ)=45$ dB assuming $\tau_{atm}=90\%$. Even if we assume pessimistically $\tau_{atm}=70\%$, $\eta_{1550nm}(90^\circ) \sim 46$~dB, indicating that $\eta_{atm}$ is only a small contribution to the total loss at zenith.}
\label{fig:losswrtelevation}
\end{figure}

The overall link efficiency (Eq.~\ref{eqn:totalloss}) combines geometry, atmospheric extinction, diffraction, optical efficiency, detector efficiency, turbulence, and pointing error. Table.~\ref{table:parameters} summarises the system parameters assumed in our model and Fig.~\ref{fig:losswrtelevation}  shows the total loss in dB with respect to angle of elevation. Elevations lower than $10^\circ$ suffer much higher losses which, along with beacon laser safety concerns and the possibility of line-of-site obstruction due to geographic topology or local environment, make them unsuitable for satellite QKD operations.

Other factors also affect the free space atmospheric channel, for example clouds, hence the additional need for real time studies of local weather conditions and historical data to estimate actual channel losses. This is further discussed in Section~\ref{sec:Cloud Cover Analysis}.

\begin{table}[htbp]
\caption{System and Model Parameters}
\label{table:parameters}%
\begin{tabular}{@{}llll@{}}
\toprule
Symbol &  Description  & Value  \\
\midrule
\(\lambda\) & Source wavelength & \(1550\) nm \\
    \(R_{e}\) & Radius of the Earth & \(6371\) km \\
    \(h\) & Altitude of the orbit & \(500\) km \\
    \(G\) & Gravitational constant & \(6.67430 \times 10^{-11}\,\text{m}^3/\text{kg}\cdot\text{s}^2\) \\
    \(M\) & Mass of the Earth & \(5.972 \times 10^{24}\,\text{kg}\) \\
    \(T_X\) & Transmitter aperture  & \(8\) cm \\
    \(R_X\) & Receiver aperture diameter & \(70\) cm \\
    \(\eta_{atm}\) & Zenith atmospheric absorption & \(0.46\) dB \\
        \(\eta_{other}\) & Lumped losses & \(20\) dB \\
\botrule
\end{tabular}
\end{table}

\section{QKD Capacity Analysis}\label{sec:QKD Capacity Analysis}

The rate of secure bits or entanglement that can be directly transferred by a lossy quantum channel is upper bounded by its secret key capacity,
\begin{equation} 
K= -\log_{2}(1 - T_{\eta}) \stackrel{\sim}{=} 1.44T_{\eta} \text{ bits/use}, 
\label{eqn:plob}
\end{equation}
where $T_{\eta}$ is the transmissivity of the channel. 
This restriction is commonly referred to as the repeaterless Pirandola-Laurenza-Ottaviani-Banchi (PLOB) bound and characterizes the fundamental rate-loss scaling for long-distance quantum optical communications in the absence of quantum repeaters~\cite{Pirandola2017}. We use the PLOB bound for comparative estimates of the capacity of the satellite-to-ground (downlink) channel to the different OGS locations~\cite{Sidhu2022,9389651,cryptography4010007}. Note that we do not include the effect of additional channel perturbations (e.g. background light, turbulence) in the current study, this can be considered in future work taking into account site-specific data such as light pollution and microclimate conditions.

In ideal cases of continuous variable (CV) and discrete variable (DV) QKD, different upper limits are set based on the type of protocol and rate-loss scaling. In the one-way switching continuous variable QKD protocol~\cite{Grosshans2003} the rate scales as $\frac{T_{\eta}}{\ln 4}$, and for the two-way protocol with coherent states and homodyne detection~\cite{Pirandola2008,Ottaviani2016} the rate scales as $\frac{T_{\eta}}{4\ln 2}$. For different discrete variable protocols, the rate scalings are: $\frac{T_{\eta}}{2}$ in Bennett and Brassard 84 (BB84)~\cite{BENNETT20147} with single-photon sources; $\frac{T_{\eta}}{2e}$ for BB84 with weak coherent pulses and decoy states~\cite{RevModPhys.81.1301}; and $\frac{T_{\eta}}{2e^{2}}$ for measurement independent QKD, where $e\approx 2.718$ is Eulers's constant~\cite{PhysRevLett.108.130502,PhysRevLett.108.130503}.

We restrict our study to the evaluation of upper bound of quantum communications using the fundamental rate loss scaling given in Eqn.~\ref{eqn:plob} irrespective of protocol and assuming asymptotic rates. With a QKD source rate of 1GHz, the secure key rate (SKR) is given by,
\begin{equation} 
SKR = K * 10^{9}  \text{ (bits/s)}.
\label{eqn:secretkeyrate}
\end{equation}
The upper bound of the secure key length (SKL) of each pass is given by the integral of key rate over duration of the overpass,
\begin{equation} 
SKL = \int_{t_{start}}^{t_{end}} SKR(t) \, dt\ 
 \text{(bits)},
\label{eqn:secretkeylength}
\end{equation}
where $t_{start,end}$ denote the times at which the satellite rises and sets below the minimum elevation limit respectively.

\begin{figure}[htbp]
\centering
\includegraphics[width=\textwidth] {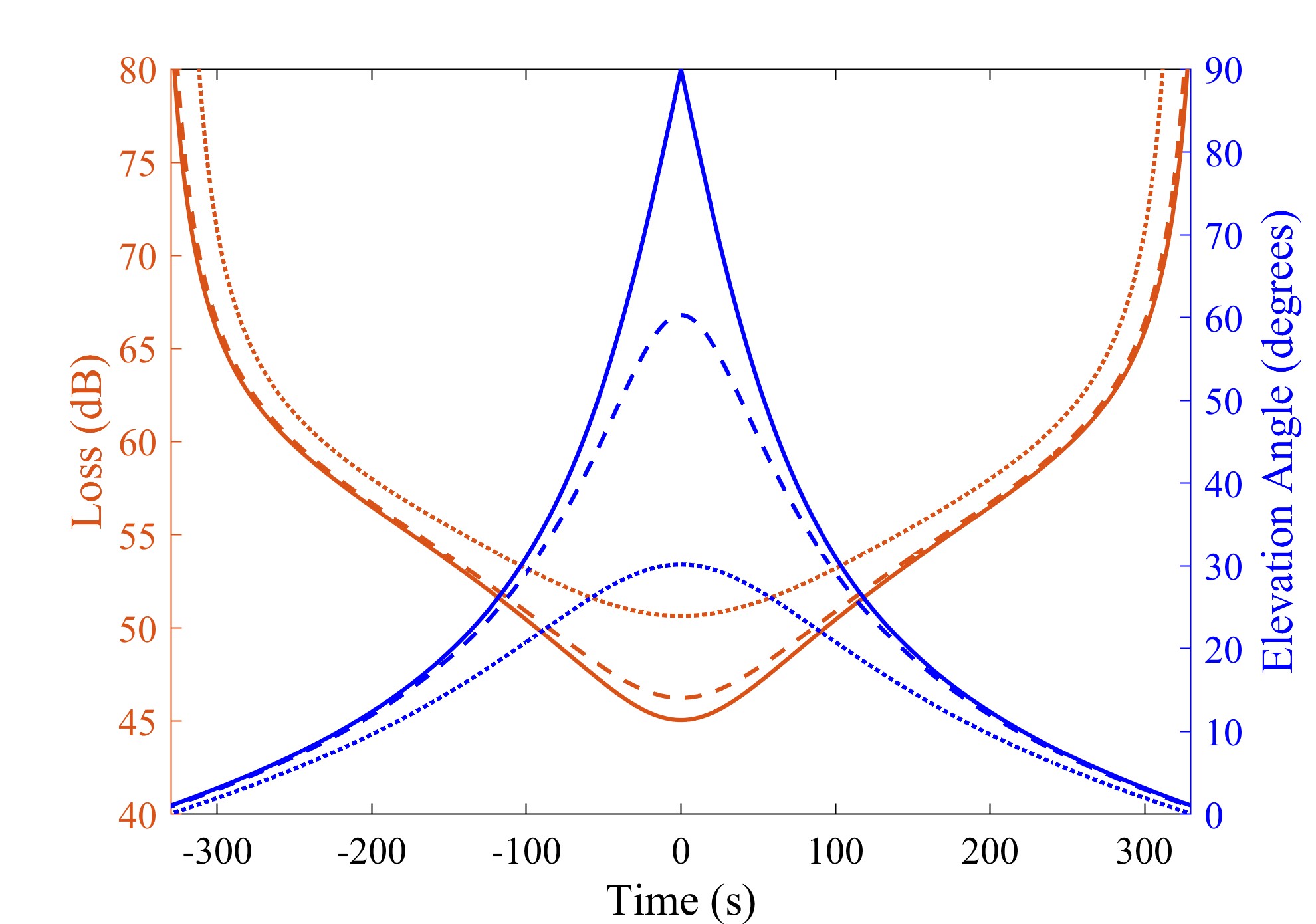} 
\caption{Overpass profiles. As a function of time (t=0 is set as time of closest approach), we plot the elevation and channel loss for $\theta_{max}=90^\circ$ (solid), $60^\circ$ (dashed), and $30^\circ$ (dotted). We assume a minimum elevation of $10^\circ$ for quantum transmission leading to contact windows of $\pm 221s$, $\pm 218s$, and $\pm 198s$ respectively.}
\label{fig:lossvstimefordifferentoverpasses}
\end{figure}

\begin{figure}[htbp]
\centering
\includegraphics[width=\textwidth] {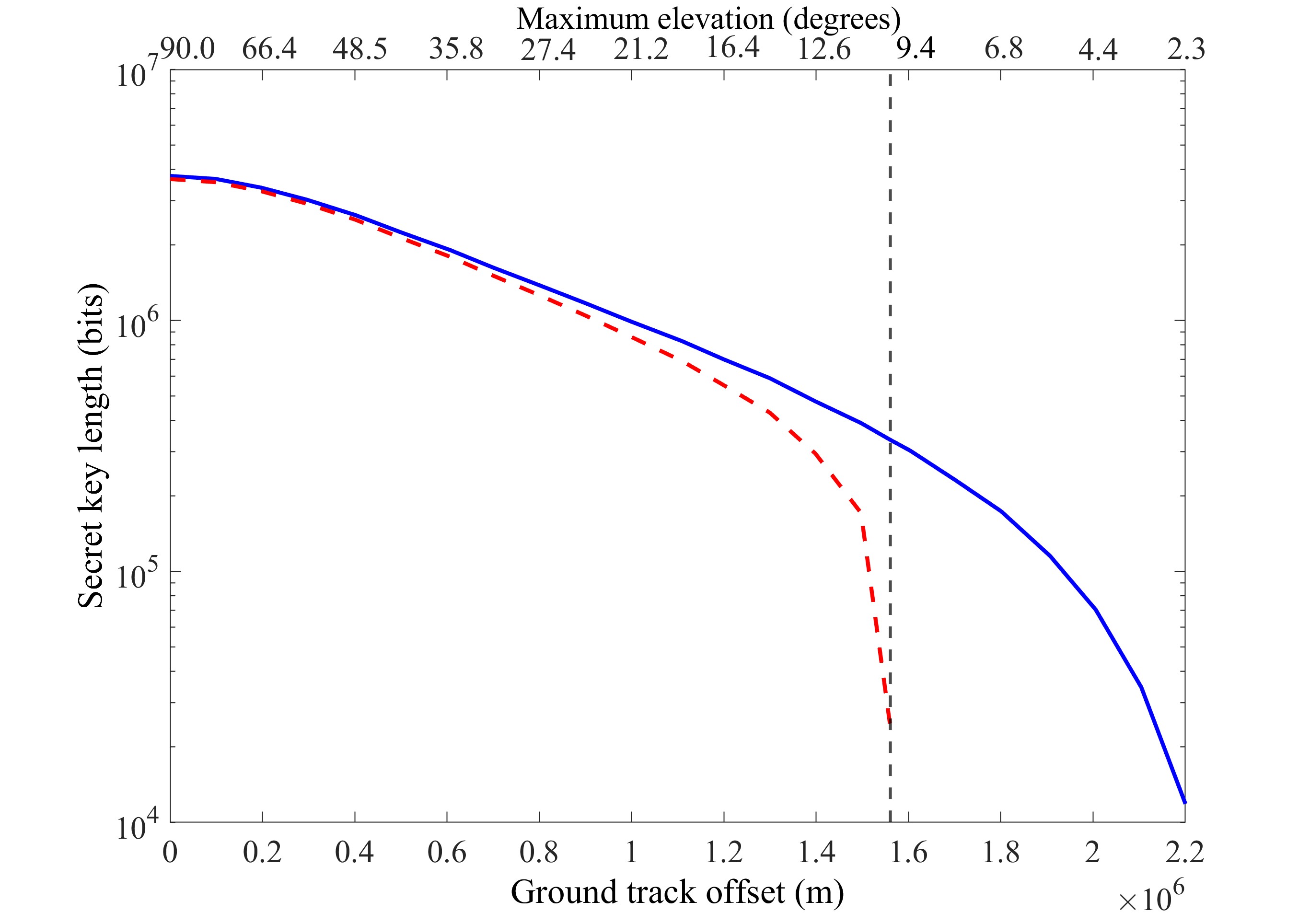} 
\caption{Secret key length Vs Ground track offset. For a fixed orbital altitude, the ground track offset and the maximum elevation angle, are equivalent. The blue solid line shows the SKL for overpasses with no elevation limit (other than the horizon) and the red dashed line shows the SKL with minimum $\theta_{max}$ set to $10^\circ$. The area under the blue curve ($0^\circ$ minimum elevation limit) is $\sim 12\%$ greater than the red dashed curve ($10^\circ$ minimum elevation limit) }
\label{fig:secretkeylengthwrtgroundtrackoffset}
\end{figure}

We calculate the SKL for passes with different $\theta_{max}$ (equivalently ground track offset $d_{min}$), shown in Fig.~\ref{fig:secretkeylengthwrtgroundtrackoffset}. The greatest ground track offset (smallest $\theta_{max}$) for which non-zero SKL is possible is denoted $d^{+}_{min}$. In general, $d^{+}_{min}$ corresponds to the ground track offset at which the satellite attains the minimum elevation limit. For a practical value of $\theta_{min}=10^\circ$ and satellite altitude $h=500\ km$, $d^{+}_{min}\approx 1500\ km$.


\subsection{Annual Secret Key Capacity}


From Fig.~\ref{fig:secretkeylengthwrtgroundtrackoffset} we can now estimate the long term average key general capacity limit of the SatQKD system, following the method in ref~\cite{Sidhu2023}. The $d_{min}$ (equally $\theta_{max}$) for each pass will vary in general due to the Earth's rotation rate and orbital period not being rationally related unless Earth synchronism is enforced~\cite{cryptography4010007}. The satellite ground track will cross the longitudinal circumference at the OGS latitude twice an orbit and over a long period $d_{min}$ will be randomly distributed in the absence of Earth synchronism.

Hence, if we consider either just the midnight or midday overpasses, the annual upper bound to SKL can be estimated by,
\begin{equation}
SKL_{year} = N_{year}\frac {SKL_{int}}{L_{lat}}\ \text{(bits)},
\label{eqn:annual secretkeylength}
\end{equation}
where $N_{\text{year}}$ is the number of orbits per year $\sim 5560$, $L_{\text{lat}}$ is the longitudinal circumference along the line of latitude at a single OGS location, and
\begin{equation}
SKL_{int} = 2 \int_{0}^{d^{+}_{min}} SKL(d_{min}) \, dd_{min} \  \text{(bit-metres)},
\label{eqn:secretkeylengthintegral}
\end{equation}
corresponding to the area under the curve in Fig.~\ref{fig:secretkeylengthwrtgroundtrackoffset}.
For $\theta_{min}=10^\circ$ (dashed red in Fig.~\ref{fig:secretkeylengthwrtgroundtrackoffset}), $SKL_{int} = 4.96 \times 10^{12}$ bit-metres and the annual secret key length for each location is given in Table~\ref{table:annualkeycapacity}. The key capacity for a single satellite covering the four locations within Ireland is similar, due to the close proximity of the geographical locations with respect to satellite coverage. The OGS altitudes are assumed to be same, with only slight differences in key capacity being due to latitude differences~\cite{Sidhu2023,islam2024finite}.

\begin{table}[htbp]
\caption{Single Satellite Clear Sky Annual Key Capacity. To simplify further analysis, we will use as a representative value the average, $1.13\times 10^9$ bits.}
\label{table:annualkeycapacity}%
\begin{tabular}{@{}llll@{}}
\toprule
Location & $L_{\text{lat}}$ ($\text{m}$)  & $M$ (bits) \\
\midrule
Dublin & $2.38 \times 10^{7}$ & $1.15 \times 10^{9}$ \\
Galway & $2.37 \times 10^{7}$ & $1.16\times 10^{9}$  \\
Cork & $2.47 \times 10^{7}$ & $1.11 \times 10^{9}$  \\
Waterford & $2.45 \times 10^{7}$ & $ 1.12 \times 10^{9}$ \\ 
\botrule
\end{tabular}
\end{table}


The above analysis does not take into account interruption due to local weather, particularly cloud cover~\cite{Polnik2020}. In the next section, we evaluate this impact and whether OGS geographic diversity can enhance single satellite performance.

\section{Cloud Cover Analysis}\label{sec:Cloud Cover Analysis}

Ideally, the satellite's transmitter and receiver at the OGS should maintain line of sight for the entire duration of each individual satellite overpass in order to maximize the amount of secure key generated. A major challenge in Ireland is the high presence of clouds, the effect of which depends on their type, amount/distribution, and altitude, along with variations in seasons and local weather conditions. Studying optical link availability in various free space optical communication scenarios is crucial due to the diverse characteristics of the propagation path, particularly in cloud-mediated environments, which significantly impact the strength of the optical link.

Various wavelength choices and optical engineering techniques have been examined in the literature to offset the effects of cloud attenuation, however the conclusion is that clouds typically limit optical communications~\cite{10.1117/12.734269}. Ref.~\cite{10487878} evaluates the implementation of the BB84 and BBM92 protocols under various losses at $1550nm$. The cloud free line of sight probability calculations are performed using integrated liquid water content statistics as inputs, to comment on the temporal and spatial correlation of clouds~\cite{7875453}. Similarly, photogrammetry and image analysis has also been used to estimate cloud distribution  and probability~\cite{rs12091382,bertin2015prediction,strathcylde_cloud}.

For the purposes of long-term cloud coverage statistics, we employ the Visual Crossing global weather database to obtain 5 years of data (January 2018 to December 2022) covering the 4 locations identified in Ireland~\cite{visual}. This type of data is widely used for Earth Observation, and climate studies providing global historic weather data records. The weather conditions for each of the locations were analyzed to find the percentage of cloud coverage on an hourly basis which we then use to estimate the modified key generation capacity~\cite{Polnik2020}. To simplify the analysis, and considering the minor differences between sites, we will use a single value $1.13\times 10^{9}$ bits to represent the baseline expected clear-sky annual key capacity, the average of Table~\ref{table:annualkeycapacity}.

We presume several essential operational capabilities for the QC system: rapid acquisition and tracking; prompt initialisation or reestablishment of quantum transmission links; quantum transmission during clear gaps between clouds; efficient downlink scheduling;  continuous accumulation of raw keys to address finite key effects~\cite{Sidhu2022,Sidhu2023}. In our study, we anticipate the system to operate even under partial cloud cover. However, increased cloud cover diminishes its efficiency, particularly in terms of critical key volume.

\subsection{Ireland cloud cover statistics}

\begin{figure}[htbp]
\centering
\includegraphics[width=\textwidth]{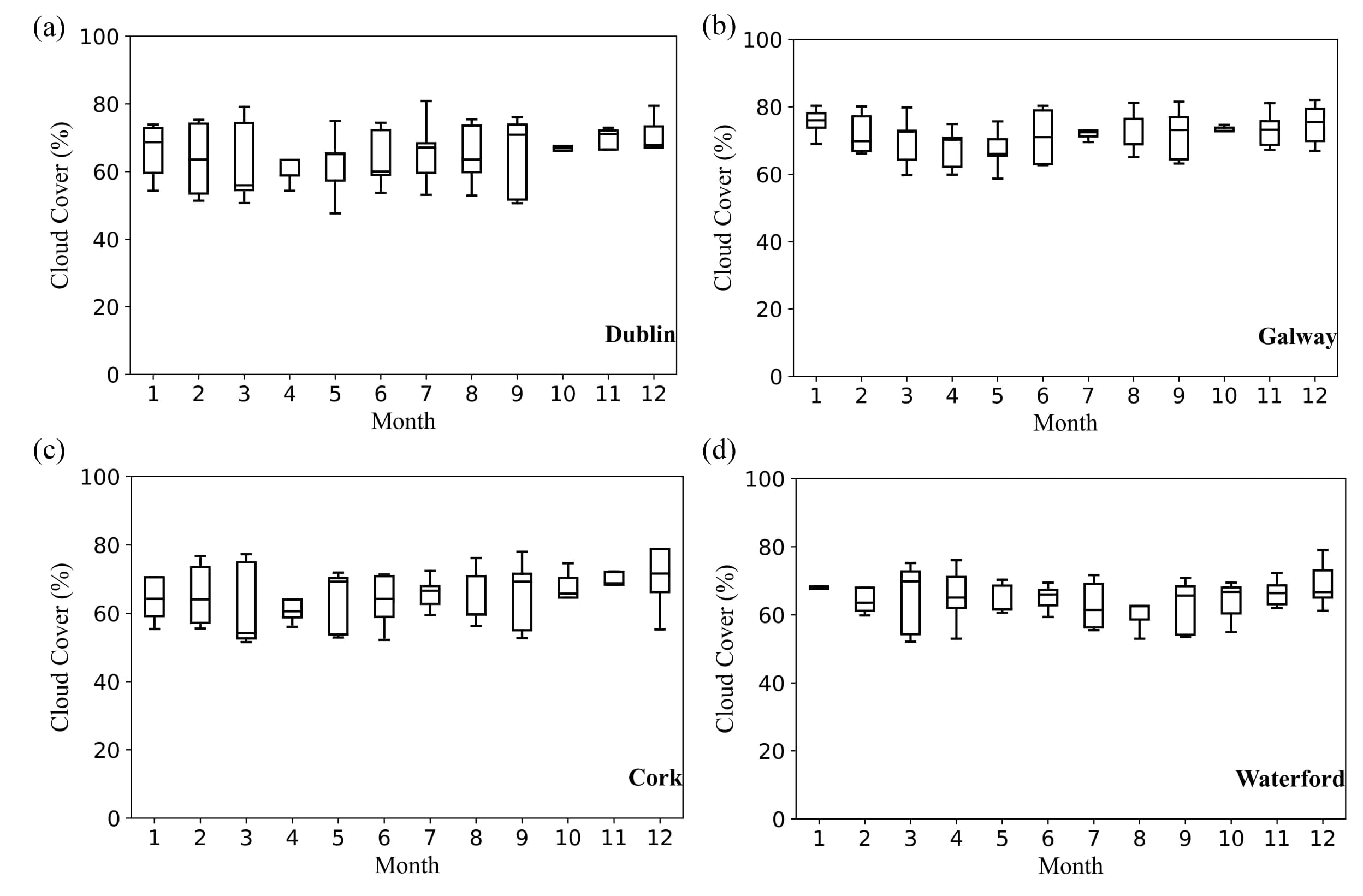} 
\caption{Cloud cover box plots of OGS sites. Each box shows the inter-quartile range and median, with whiskers indicating the total range of cloud cover within each month. Data coverage is January 2018 to December 2022. More consistently low cloud coverage periods can be identified at each location: Dublin in April (4.6\% variability and 58.84\% cloud cover); Galway in May (4.9\% variability and 66.04\% cloud cover); Cork in April (5.1\% variability and 60.5\% cloud cover); and Waterford in August (3.9\% variability and 62.5\% cloud cover).
}
\label{fig:monthlycloud}
\end{figure}

We first perform an initial cloud cover analysis, examining seasonal and hourly variation. The monthly averaged cloud cover across the year is shown in Fig.~\ref{fig:monthlycloud} as box charts with the median and variability in cloud cover at each site. The overall lowest median cloud cover was in Cork during March (54.13\%), while the highest median cloud cover was in Galway in January (75.97\%). However, it is also desirable for low variability in cloud cover to provide greater consistency in site performance.

\begin{figure}[htbp]
\centering
\includegraphics[width=\textwidth]{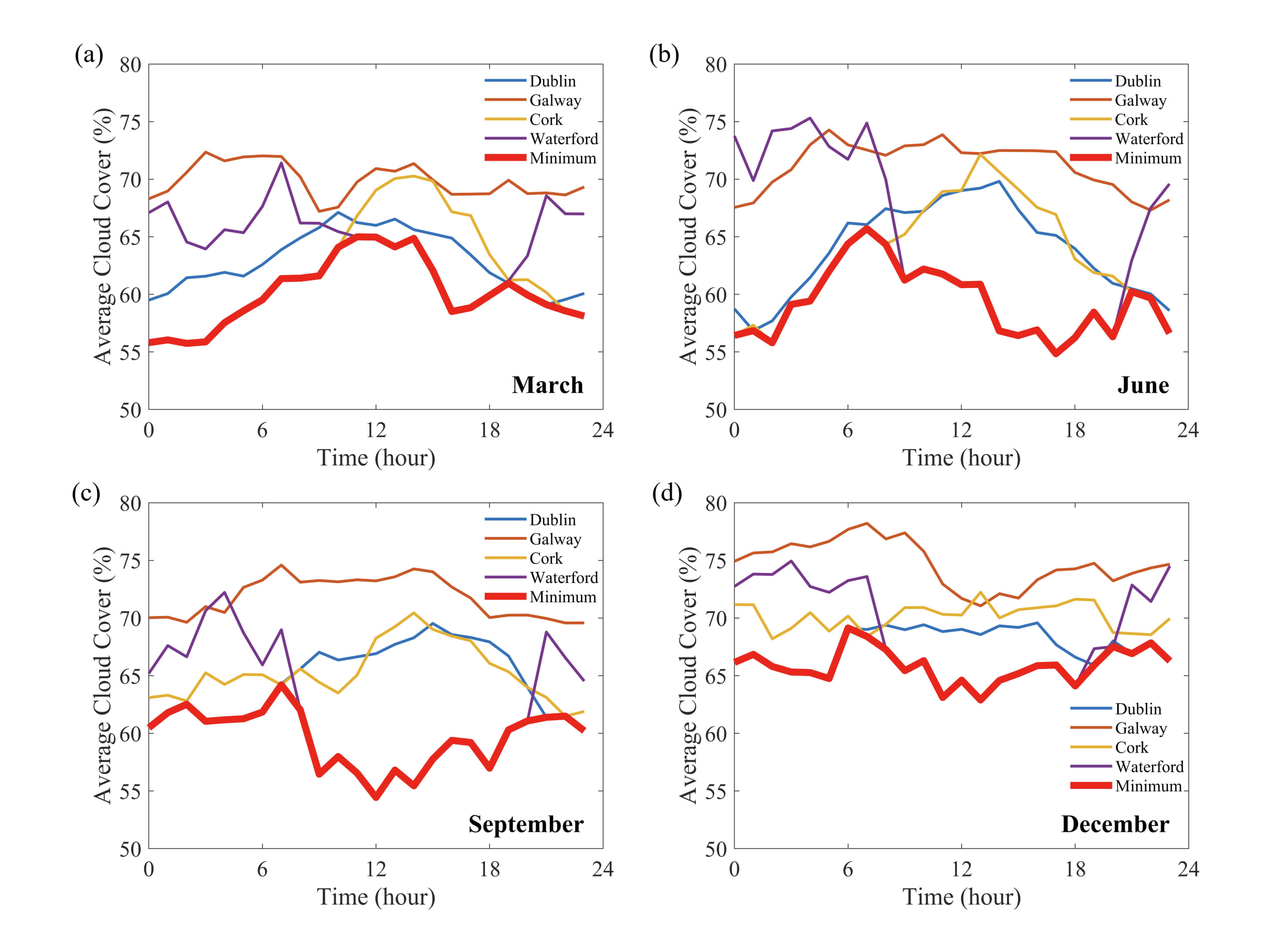} 
\caption{Seasonal analysis of cloud cover. Average cloud cover and minimum (red solid line) value at each hour for months (a) March, 
(b) June, (c) September and (d) December, from cloud cover data (2018 - 2022) of four locations in 
Ireland. The red solid line shows lowest cloud cover at each
hour available from four locations.}\label{fig:hourly_MarJunSepDec}
\end{figure}
\begin{figure}[tbh]
\centering
\includegraphics[width=\textwidth] {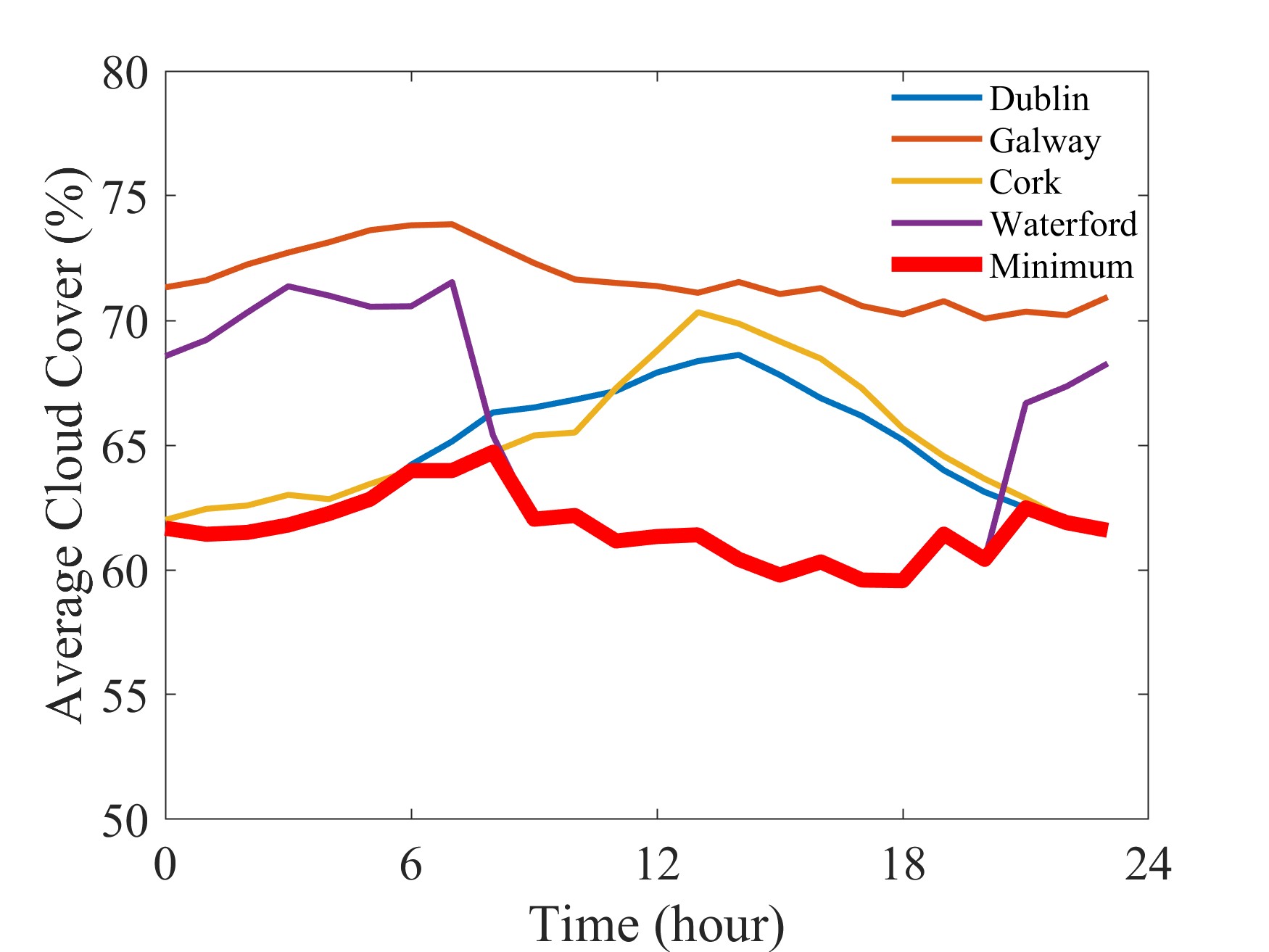} 
\caption{Annual analyis of cloud cover. Average cloud cover and minimum (red solid line) at each hour at different locations from 
cloud cover data of all months from 2018 to 2022. The red solid line shows lowest cloud cover at each
hour available from four locations.}
\label{fig:hourly}
\end{figure}

\begin{figure}[htbp]
\centering
\includegraphics[width=\textwidth]{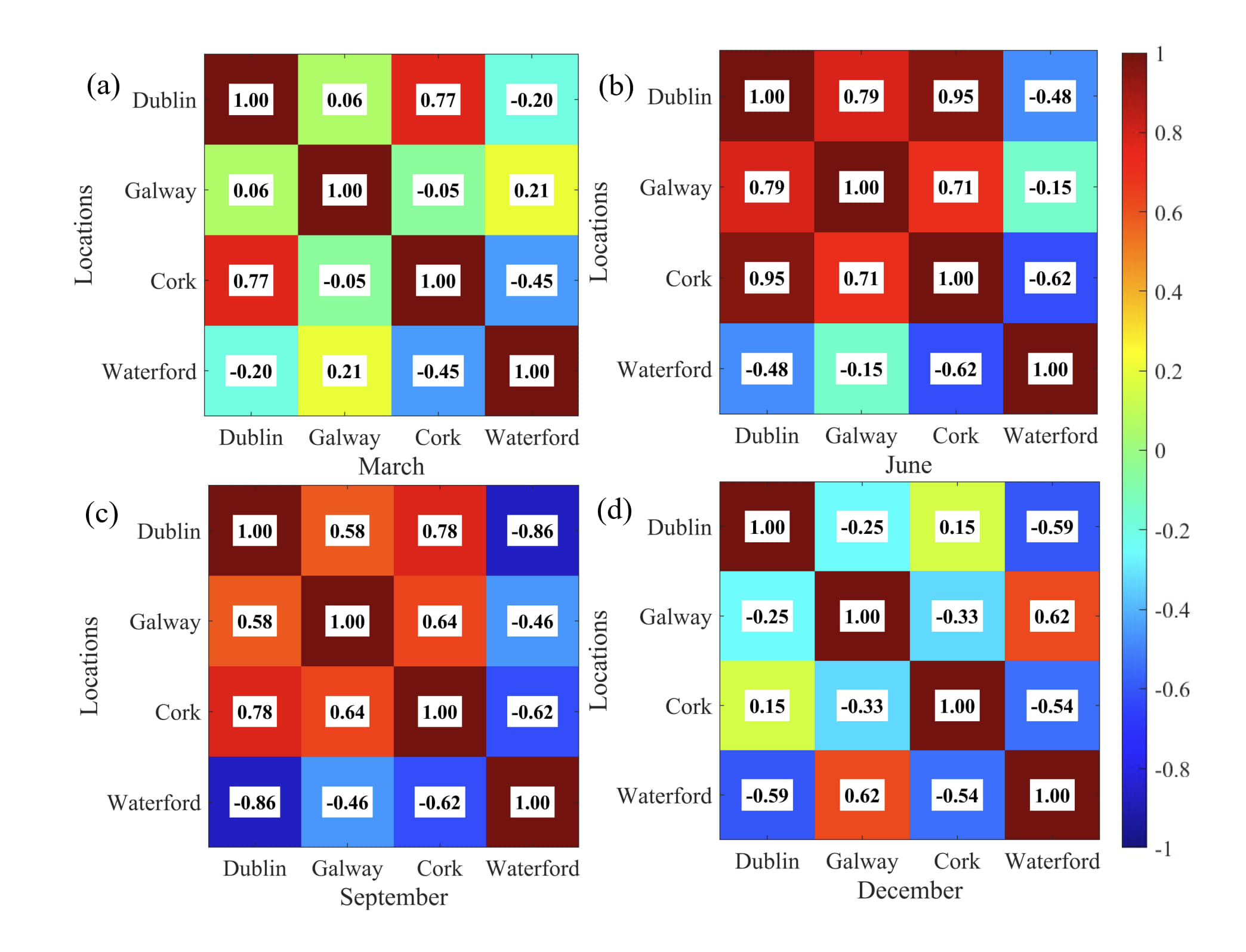} 
\caption{Cross correlation of seasonal cloud cover average. Correlation of average cloud cover percentage of different OGS sites in (a) March, (b) June, (c) September, and (d) 
December. In the months of June and September, Dublin and Cork are correlated, whereas Waterford is anti-correlated. In the month of December, Dublin and Waterford maintain anti-correlation, while Cork is uncorrelated.}\label{fig:crosscorrelation_MarJunSepDec}
\end{figure}

If we consider each site, then we can identify the time of day for which cloud cover is at a minimum on average. Fig.~\ref{fig:hourly_MarJunSepDec} shows the average hourly variation for months representative of different seasons, and Fig.~\ref{fig:hourly} shows the annual average for each hour. On both figures, the minimum across all 4 sites is also indicated. We observe high variability throughout the day and across sites, e.g. in Fig.~\ref{fig:crosscorrelation_MarJunSepDec} Waterford in June and September shows much lower cloud cover during the day than the other 3 sites. Hence, this motivates a closer examination of the inter-site variations for determining optimal OGS diversity.

\begin{figure}[tbh]
\centering
\includegraphics[width=\textwidth] {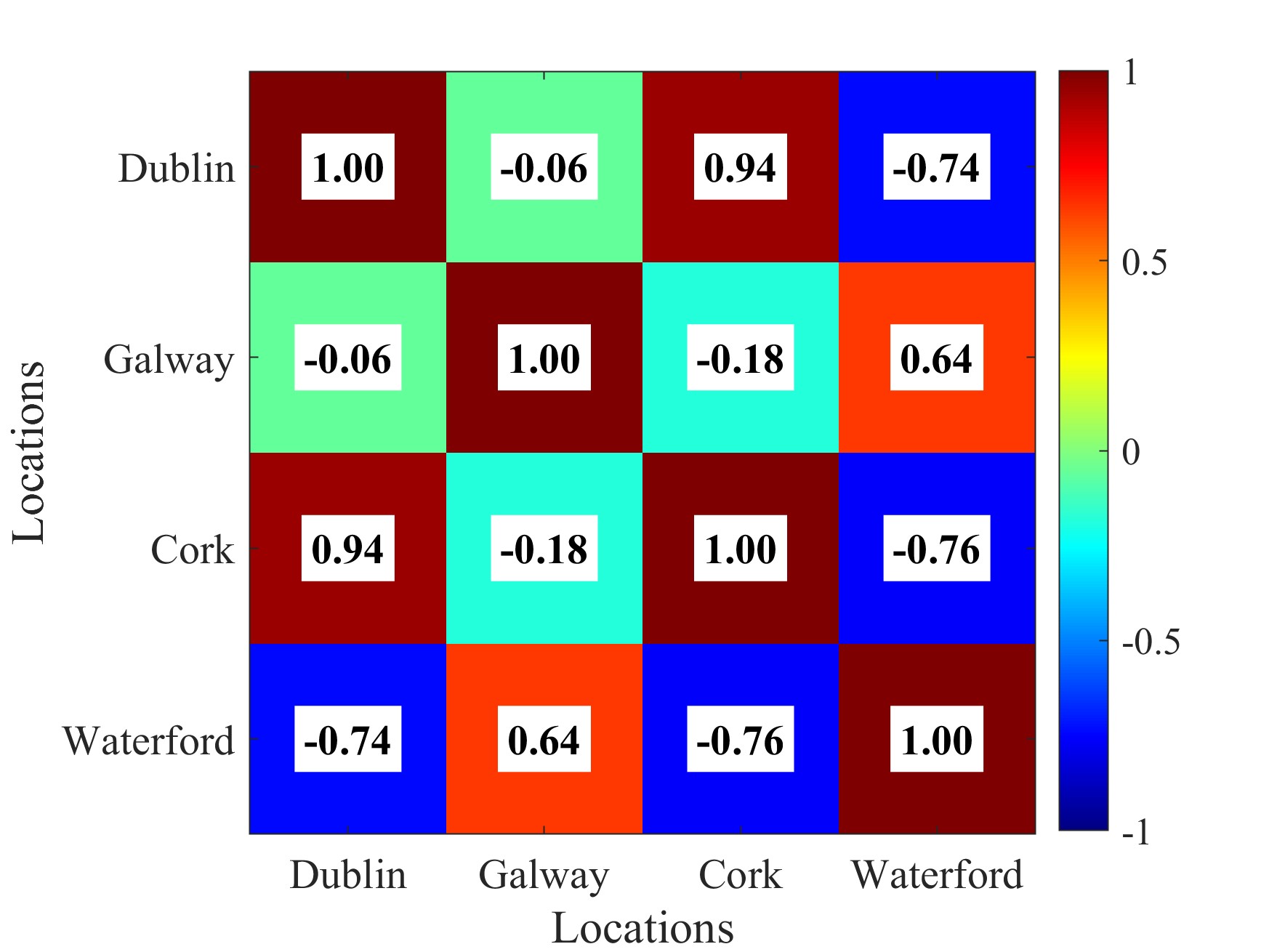} 
\caption{Cross correlation of cloud cover percentage from data of all months between 2018-2022. Positive correlation shows similar cloud cover pattern and negative shows anti correlated cloud cover pattern and value close to zero shows that sites are uncorrelated. Dublin and Cork shows strong positive correlation where as Waterford is anti-correlated with them.}
\label{fig:crosscorrelation}
\end{figure}

The cross correlation of cloud cover between locations (Fig.~\ref{fig:crosscorrelation_MarJunSepDec}) for June and September supports the above observation. Dublin and Cork are correlated, whereas Waterford is anti-correlated. In December, Dublin and Waterford maintain anti-correlation, while Cork is uncorrelated. The cross-correlation across the whole year (Fig.~\ref{fig:crosscorrelation}) shows strong correlation between Dublin and Cork, indicating similar cloud cover patterns. Dublin and Waterford (-0.73997), as well as Cork and Waterford (-0.75846), show strong anti-correlations, indicating inverse cloud cover patterns between these locations. This pattern in cloud cover correlations suggests that strategically using OGSs at different sites can significantly enhance the overall lowest cloud coverage possible to optimize satellite communication opportunities, maximising the chances that at least one ground station is available for an efficient link. A combination of OGSs at Dublin, Cork, and Waterford is likely to result in improved cloud cover compared to a single site. In the following section, we discuss more explicit modeling of cloud cover with respect to various OGS combinations, examining the improvement in key capacity that ground station diversity may provide.

\subsection{Single satellite - OGS Network Availability}

Here, we consider an OGS network consisting of multiple ground stations that establish secure keys with a single satellite. We assume that the satellite can communicate with a single OGS in each pass. For simplicity, we do not model an explicit orbit, but work with average key generation rates which should be sufficient for determining long-term trends. For our initial analysis, we consider midnight and midday overpasses, assuming daylight operational capability in the latter case.

\begin{figure}[htbp]
    \centering
    \includegraphics[width=\textwidth]{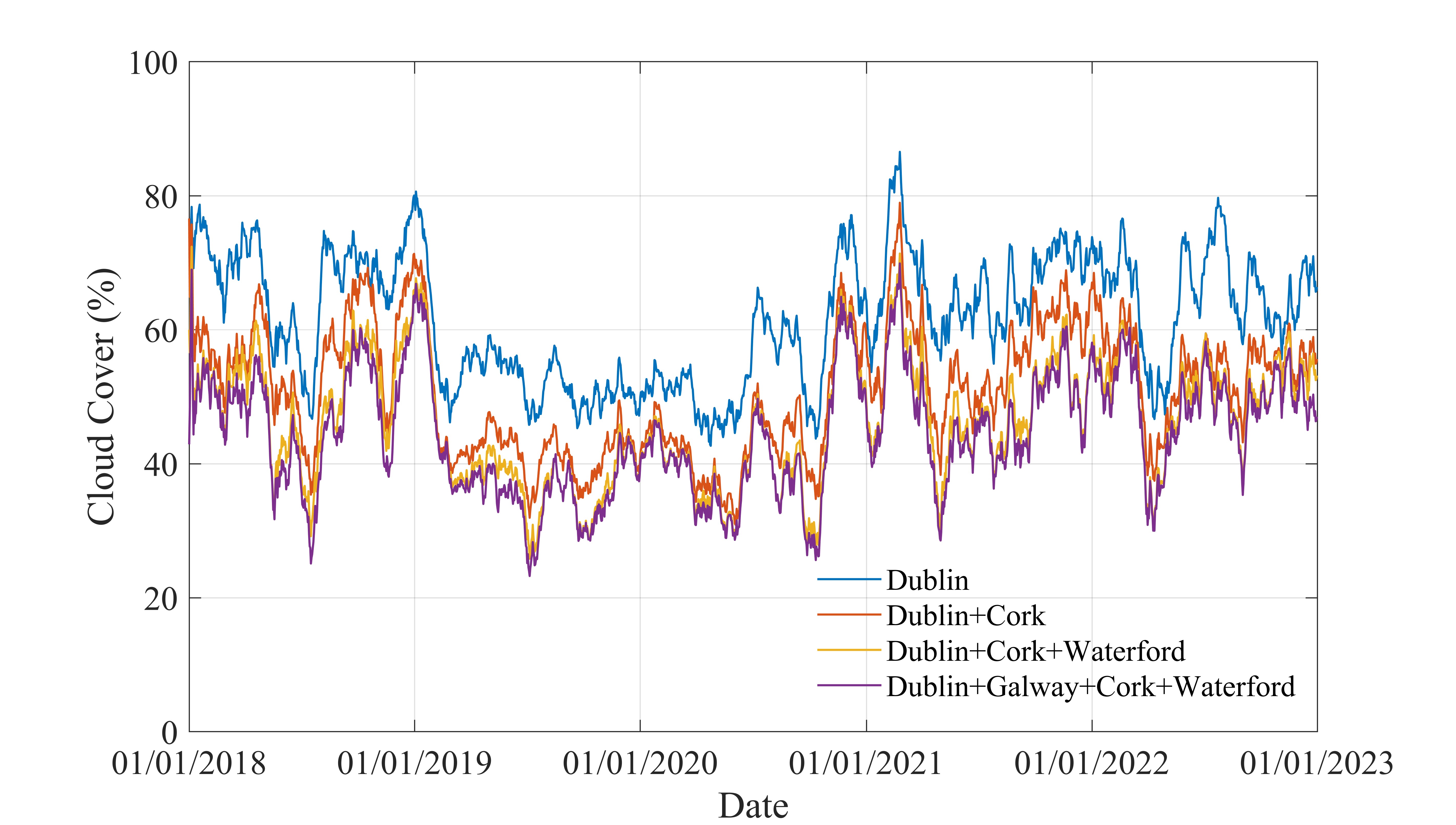}
    \caption{Midnight 30-day Running Average Cloud Cover. We plot the 30 day running average minimum cloud cover for a combination of sites over the period of the sample weather data. A noticeable improvement is seen when going from a single OGS site in Dublin, to 2 sites in Dublin and Cork. There is a significant difference between the years 2019 - 2020 compared with 2018 and 2021 - 2022.}
    \label{fig:midnightaverage}
\end{figure}

\begin{figure}[htbp]
    \centering
    \includegraphics[width=\textwidth]{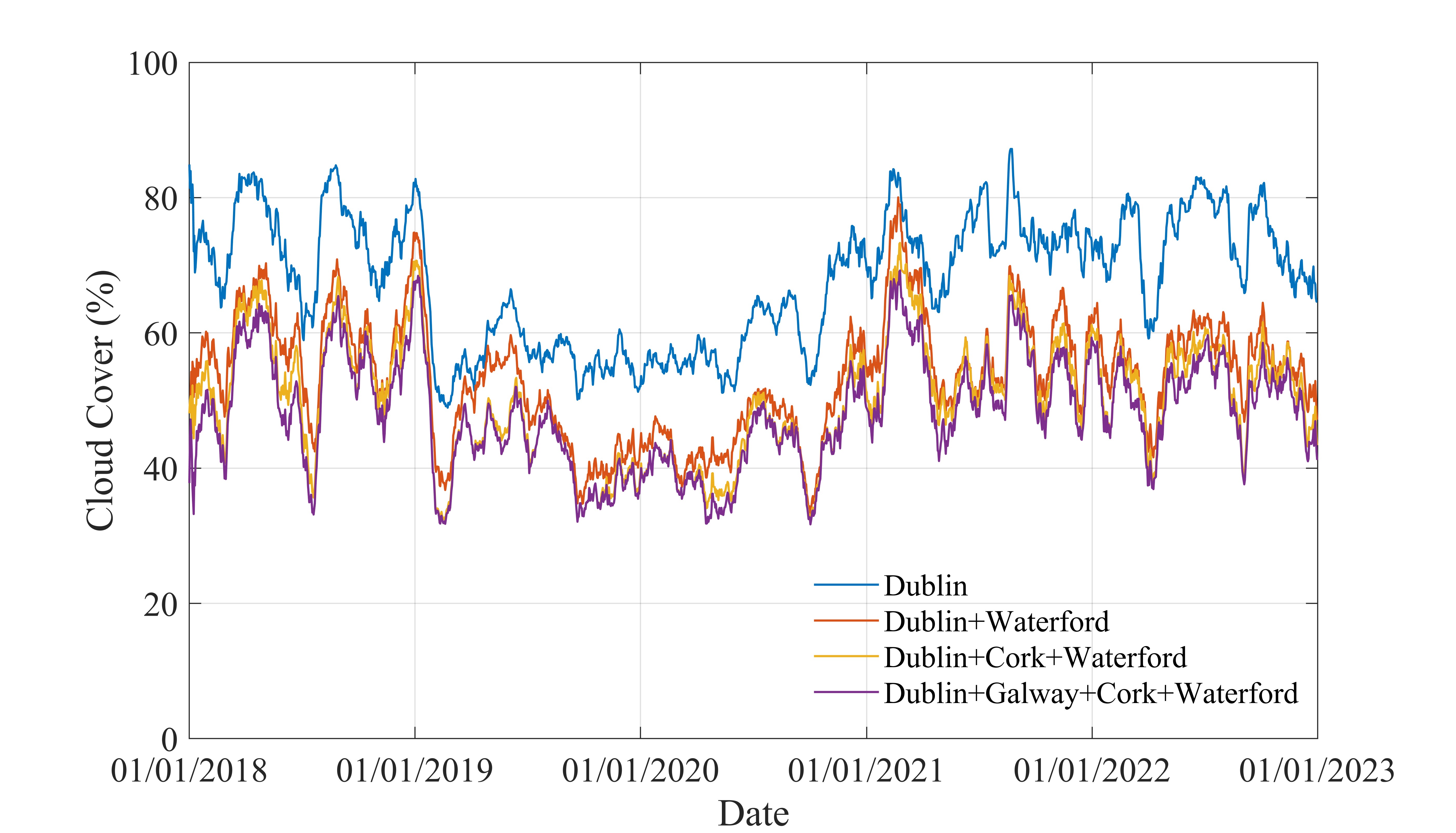}
    \caption{Midday 30-day Running Average Cloud Cover. We plot the 30 day running average minimum cloud cover for a combination of sites over the period of the sample weather data. A noticeable improvement is seen when going from a single OGS site in Dublin, to 2 sites in Dublin and Waterford. 
    }
    \label{fig:middayaverage}
\end{figure}

For the data for each day at the overpass times considered, the cloud cover at each site is analysed based on different site combinations reflecting the choice of 1, 2, 3, or 4 OGS locations (15 cases). The site within each combination with the lowest observed cloud cover is chosen for downlink by the satellite. The selected sites' cloud cover over the entire 5 years is then averaged to estimate the scaled secure key length that could be generated by a single satellite with an OGS network consisting of that combination of sites. We summarise the analysis for midnight (Table~\ref{table:locationmidnightcounts}) and midday (Table~\ref{table:locationmiddaycounts}) overpass times averaged over the entire 5 years data set, as well as 30-day running averages (Fig.~\ref{fig:midnightaverage} and Fig.~\ref{fig:middayaverage} respectively).

We discuss 15 cases of OGS combinations: 4 cases of a single OGS site; 6 cases of dual OGS sites, 4 cases of three-OGS sites; and finally 1 case of all four OGS sites. In case of midnight overpass analysis (Table~\ref{table:locationmidnightcounts}), combining Dublin and Cork results in a mean minimum cloud cover of 51.1\%, a significant improvement over the 61.7\% when using only Dublin. Adding Waterford to the Dublin and Cork combination reduces the mean minimum cloud cover to 46.1\%. While all four sites are taken into account (Dublin, Galway, Cork, and Waterford) brings the mean down to 44.4\%, the incremental improvement is relatively minor compared to the three-site combination. In the case of midday overpass analysis (Table~\ref{table:locationmiddaycounts}), using only the Waterford site results in a mean minimum cloud cover of 61.3\%. When combining Dublin and Waterford, the mean minimum cloud cover significantly improves to 53.5\%. Further adding Cork to the Dublin and Waterford combination reduces the mean minimum cloud cover to 49.9\%. As for midnight passes, including all four sites (Dublin, Galway, Cork, and Waterford) results in a minor improvement to 48.1\%. Overall, the addition of multiple sites mitigates the impact of cloud cover but with diminishing benefits.

\begin{table}[htbp]
    \centering
    \begin{tabular}{|l|rrrr|r|}
    \hline
    \textbf{OGS Combinations} & \multicolumn{4}{|c|}{\textbf{Min Cloud Cover Number}} & \textbf{Cloud Cover} \\
     & Dublin & Galway & Cork & Waterford & Mean Min \%\\
    \hline 
        Dublin & 1826 &  &  &  & \textbf{61.7}\\
        Galway &  & 1826 &  &  & 71.3\\
        Cork &  &  & 1826 &  & 62.0\\
        Waterford &  &  &  & 1826 & 68.6\\
        \hline
        Dublin+Galway & 1281 & 545 &  &  & 56.3\\
        Dublin+Cork & 925 &  & 901 &  & \textbf{51.1}\\
        Dublin+Waterford & 1183 & & & 643 & 52.8\\
        Galway+Cork & & 596 & 1230 & & 56.4\\
        Galway+Waterford & & 960 & & 866 & 57.9\\
        Cork+Waterford & & & 1185 & 641 & 52.9\\
        \hline
        Dublin+Galway+Cork & 793 & 272 & 761 & & 49.0\\
        Dublin+Galway+Waterford & 908 & 378 & & 540 & 49.3\\
        Dublin+Cork+Waterford & 695 & & 675 & 456 & \textbf{46.1}\\
        Galway+Cork+Waterford & & 421 & 856 & 549 & 49.4\\
        \hline
        Dublin+Galway+Cork+Waterford & 601 & 226 & 566 & 433 & \textbf{44.4}\\
        \hline
    \end{tabular}
    \caption{OGS Site Combination Minimum Cloud Cover. For each combination of OGS locations, we tally the number of times that a location has the minimum cloud cover at 00:00 over the 5 year period (1826 nights), together with the average minimum cloud cover. The best choices for 1, 2, and 3 sites are: Dublin (61.7\%); Dublin+Cork (51.1\%); Dublin+Cork+Waterford (46.1\%). Choosing all 4 sites (44.4\%) only results in a minor incremental improvement.}
    \label{table:locationmidnightcounts}
\end{table}

\begin{table}[htbp]
    \centering
    \begin{tabular}{|l|rrrr|r|}
    \hline
    \textbf{OGS Combinations} & \multicolumn{4}{|c|}{\textbf{Min Cloud Cover Number}} & \textbf{Cloud Cover} \\
     & Dublin & Galway & Cork & Waterford & Mean Min \%\\
    \hline 
        Dublin & 1826 &  &  &  & 67.9\\
        Galway &  & 1826 &  &  & 71.4\\
        Cork &  &  & 1826 &  & 68.8\\
        Waterford &  &  &  & 1826 & \textbf{61.3}\\
        \hline
        Dublin+Galway & 1146 & 680 &  &  & 61.4\\
        Dublin+Cork & 911 &  & 915 &  & 59.6\\
        Dublin+Waterford & 669 & & & 1157 & \textbf{53.5}\\
        Galway+Cork & & 793 & 1033 & & 61.6\\
        Galway+Waterford & & 657 & & 1169 & 54.8\\
        Cork+Waterford & & & 649 & 1177 & 53.9\\
        \hline
        Dublin+Galway+Cork & 701 & 435 & 690 & & 56.2\\
        Dublin+Galway+Waterford & 515 & 347 & & 540 & 50.7\\
        Dublin+Cork+Waterford & 444 & & 444 & 938 & \textbf{49.9}\\
        Galway+Cork+Waterford & & 416 & 457 & 953 & 50.8\\
        \hline
        Dublin+Galway+Cork+Waterford & 376 & 263 & 362 & 825 & \textbf{48.1}\\
        \hline
    \end{tabular}
    \caption{OGS Site Combination Minimum Cloud Cover. For each combination of OGS locations, we tally the number of times that a location has the minimum cloud cover at 12:00 over the 5 year period (1826 days), together with the average minimum cloud cover. The best choices for 1, 2, and 3 sites are: Waterford (61.3\%); Dublin+Waterford (53.5\%); Dublin+Cork+Waterford (49.9\%). Choosing all 4 sites (48.1\%) only results in a minor incremental improvement.}
    \label{table:locationmiddaycounts}
\end{table}

\begin{table}[htbp]
\caption{Single-Satellite Cloud-cover Weighted Annual Key Capacity for Midnight (Case I) and Midday (Case II) Overpasses.}
\label{table:average_annual_keycapacity}
\begin{tabular}{@{}llll@{}}
\toprule
Location   & Case I (bits) & Case II (bits)  \\
\midrule
Dublin  & $0.43 \times 10^{9}$ & $0.44 \times 10^{9}$ \\
Dublin+Cork  & $0.56\times 10^{9}$ & $0.45 \times 10^{9}$ \\
Dublin+Waterford  & $0.53\times 10^{9}$ & $0.53\times 10^{9}$ \\
Dublin+Cork+Waterford & $0.61 \times 10^{9}$ & $0.57 \times 10^{9}$ \\
Dublin+Cork+Galway+Waterford& $ 0.63 \times 10^{9}$ & $ 0.59 \times 10^{9}$ \\ 
\botrule
\end{tabular}
\begin{flushleft}
\textbf{Case I:} Midnight OGS combinations key capacity\\
\textbf{Case II:} Midday OGS combinations key capacity \\
\end{flushleft}
\end{table}

\section{Conclusion}\label{sec:Conclusion}

We have provided quantification of the impact of cloud cover on SatQKD for Ireland that, in particular, suffers from significant weather effects due to predominantly Westerly wind patterns across the Atlantic. Out of the 4 identified potential OGS locations, the best single site availability was 38.3\% (38.7\%) at midnight (midday) over the period 2018-2022, which reduces the secure key capacity to approximately a third of the ideal value based on the asymptotic PLOB bound. However, by exploiting anti-correlation in cloud cover, a single satellite can achieve greater availability, up to 55.6\% (51.9\%) across all 4 locations, though even only with 2 OGSs a significant improvement is possible. A preliminary 2-site OGS network would consist of either Dublin and Cork or Dublin and Waterford, depending on whether midnight (48.9\%) or midday (46.5\%) availability were prioritised. We should note that the availability of the combination of Cork and Waterford is close to the preceding choices (midnight 47.1\% and midday 46.1\%). 

Overall, the choice of Dublin and Waterford would be a reasonable compromise balancing midday and midnight availability together with other factors such as density of potential users. This combination should provide a PLOB bounded secure key capacity of half a gigabit for the modestly specified SatQKD system considered here. For a WCP decoy-state BB84 implementation, this gives 250,000 AES-256 keys per year for either midday or midnight passes, without taking into account system imperfections, noise, and error correction inefficiency. 


Locating clear skies as often as possible at OGS locations is the most significant condition for QC. More detailed studies are recommended as a follow-up based on real time on-site data records including background light levels, turbulence, and other microclimate effects. Our model assumed sufficient raw key accumulation to avoid finite size effects~\cite{Sidhu2022,Sidhu2023} and large key buffers to cover long periods of OGS unavailability. However, the current approach provides an initial estimate of the feasibility of satellite QKD in Ireland and the effectiveness of OGS site diversity as a mitigation against local cloud cover and is an important step in Ireland for the development of the future quantum internet.

\backmatter

\bmhead{Abbreviations}
AGT, Atmospheric Generator Toolkit;
BB84 protocol, Bennett and Brassard 1984 protocol;
BBM92 protocol, Bennett, Brassard and Mermin 1992 protocol;
CFLOS, Cloud Free Line of Sight;
{EuroQCI}: European Quantum Communication Infrastructure;
IRIS$^{2}$, Infrastructure for Resilience, Interconnectivity and Security by Satellite;
LEO: Low Earth Orbit;
MODTRAN, MODerate resolution atmospheric TRANsmission;
OGS, Optical Ground Station;
OPS-SAT VOLT, Optical Scylight Satellite Versatile Optical Lab for Telecommunications;
PLOB bound, Pirandola-Laurenza-Ottaviani-Banchi bound;
QC, Quantum Communications;
QEYSSat, Quantum Encryption and Science Satellite;
QKD, Quantum Key Distribution;
QuNET, Quantum Network;
SAGA, Security And cryptoGrAphic mission;
SatQKD, Satellite Quantum Key Distribution;
SKL, Secure Key Length; 
SKR, Secure Key Rate;
SPOQC, Satellite Platform for Optical Quantum Communications;
SSO, Sun Synchronous Orbit;
TLE, Two Line Element.

\bmhead{Acknowledgements}
Not applicable.

\bmhead{Author contributions}
The simulations and data analysis were performed by NLA. The effort was conceived and supervised by JH, DKLO, and DK. NLA and DKLO wrote the draft and all authors reviewed the manuscript.

\bmhead{Funding}
This work is supported by South East Technological University postgraduate research programme and in part by a Grant from Science Foundation Ireland under Grant number 13/RC/2077\_{P2} and Grant number 21/US-C2C/3750. DKLO acknowledges support from the EPSRC Quantum Technology Hub in Quantum Communication (EP/T001011/1). This work was supported by the EPSRC International Network in Space Quantum Technologies INSQT (EP/W027011/1).

\bmhead{Data availability}
The data sets used were commercially obtained from Visual Crossing~\cite{visual}.


\section*{Declarations}

\bmhead{Ethics approval and consent to participate}
Not applicable.

\bmhead{Competing interests}
The authors declare no competing interests.

\bmhead{Consent for publication}
Not applicable.









\bibliography{IrelandSatQKDWeather_EPJ-QT_Submit}

\end{document}